\documentclass[manuscript]{aastex} 
\newcommand{\vhbb}{$\Delta V_{\rm HB}^{\rm Bump}\,$}

\newcommand{\Iso}[2]{^{#1}{\mathrm{#2}}}   
  
\begin{document}   
   
\title{Red Giant Branch stars: the theoretical framework.}   
   
\author{Maurizio Salaris}   
\affil{Astrophysics Research Institute, Liverpool John Moores    
       University, Twelve Quays House, Egerton Wharf, Birkenhead CH41    
       1LD, United Kingdom; ms@astro.livjm.ac.uk}   
\author{Santi Cassisi}   
\affil{Osservatorio Astronomico di Collurania,   
 via M. Maggini, 64100 Teramo, Italy; cassisi@te.astro.it}   
\author{Achim Weiss}   
\affil{Max Planck Institut f\"ur Astrophysik,   
Karl-Schwarzschild-Str. 1, 85748 Garching, Germany; weiss@mpa-garching.mpg.de}

\received{}   
\accepted{}

\pagebreak    
\begin{abstract}   
 
Theoretical predictions of Red Giant Branch stars' effective temperatures, 
colors, luminosities and surface chemical abundances are a necessary tool for the 
astrophysical interpretation of the visible--near infrared integrated 
light from unresolved stellar populations, the Color-Magnitude-Diagrams 
of resolved stellar clusters and galaxies, and spectroscopic 
determinations of red giant chemical abundances. 
On the other hand, the comparison with empirical constraints provides a 
stringent test for the accuracy of present generations of red giant models.
 
We review the current status of red giant stars' modelling, discussing 
in detail the still existing uncertainties affecting the model input 
physics (e.g., electron conduction opacity, treatment of the superadiabatic convection), 
and the adequacy of the physical assumptions built into the 
model computations.
  
We compare theory with several observational
features of the Red Giant Branch in galactic globular clusters, such as the luminosity
function \lq{bump}\rq, the luminosity of the Red Giant Branch tip 
and the envelope chemical
abundance patterns, to show the level of agreement between current  
stellar models and empirical data concerning the stellar luminosities, 
star counts, and surface chemical abundances.
     
\end{abstract}   
   
\keywords{atomic processes -- convection -- globular clusters: general -- 
nuclear reactions, nucleosynthesis, 
abundances -- stars: atmospheres -- stars: evolution -- stars: interiors -- 
stars: Hertzsprung-Russell (HR) and C-M diagrams}   
   
\pagebreak    
\section{Introduction}   
   
The Red Giant Branch (RGB) is one of the most prominent   
and well populated features in the Color-Magnitude-Diagram (CMD) of   
stellar populations with ages larger than about $1.5 - 2$ Gyr.    
The theoretical modelling of RGB stars plays therefore a wide ranging role,   
involving various fields of galactic and extragalactic astrophysics.   
   
Since RGB stars are cool, reach high luminosities during    
their evolution, and their evolutionary timescales are relatively long,   
they provide a major contribution to the integrated   
bolometric magnitude and to integrated colors and spectra   
at wavelengths larger than about 900 -- 1000 nm of old   
distant, unresolved stellar populations (e.g. Renzini \& Fusi-Pecci~1988;    
Worthey~1994).   
A correct theoretical prediction of the RGB spectral properties and colors  
is thus of paramount importance for   
interpreting observations of distant stellar clusters and galaxies   
using population synthesis methods, but also for    
determining the ages of resolved globular and open clusters, by means of isochrone   
fitting techniques.   
  
The $I$-band brightness of the tip of the RGB (TRGB) provides a     
robust standard candle, very much independent of the stellar age and initial   
chemical composition, which can allow to obtain  
reliable distances out to about 10 Mpc using $HST$ observations   
(e.g., Lee, Freedman \& Madore~1993 -- LFM93). Due to the lingering   
uncertainties on the empirical determination of the TRGB brightness   
zero point, RGB models provide an independent path to the   
calibration of this important standard candle   
(Salaris \& Cassisi~1997). Also the theoretical calibration of   
the luminosity of Horizontal   
Branch (HB) stars and their RR Lyrae population    
(whose parallax-based distances still show a very large error bar;   
see, e.g., Groenewegen \& Salaris~1999) is dependent   
on the correct modelling of the previous RGB phase, since    
HB luminosities (like the TRGB ones)    
are determined by the value of the electron degenerate    
He-core mass ($M_{core}^{He}$) at the end of the RGB evolution.   
   
Predicted evolutionary timescales along the RGB phase play a    
fundamental role in the determination of the initial He abundance 
of globular cluster stars through the R parameter    
(number ratio between HB stars and RGB stars   
brighter than the HB at the RR Lyrae instability strip level; see, e.g.   
Iben~1968a, Sandquist~2000, Zoccali et al.~2000), while   
an accurate modelling of the mixing mechanisms efficient  
in the RGB stars is necessary to correctly  
interpret spectroscopic observations of their surface chemical   
abundance patterns.    
  
These various applications of RGB stellar models to fundamental    
astrophysical   
problems crucially rely on the ability of theory to predict correctly:   
   
\noindent   
-- the CMD location (in $T_{\rm eff}$ and color) and extension    
(in brightness) of the RGB as a function of the initial chemical composition   
and age;   
   
\noindent   
-- the evolutionary timescales (hence the relative numbers of   
stars at different luminosities) all along the RGB;   
   
\noindent   
-- the RGB stars' physical and chemical structure, as well as   
their evolution with time.   
   
The main goal of this review is to discuss the existing uncertainties  
in theoretical RGB models, and to assess the reliability of their predictions.  
  
In \S 2 we present an outline of RGB stellar evolution,  
while in \S 3 and \S 4 fundamental properties of    
the CMD and luminosity functions of RGB models are reviewed.   
\S 5 analyzes the input physics currently used for computing RGB models,   
and how related uncertainties affect the outcome of the calculations;   
\S 6, \S 7 and \S 8 discuss observational tests for the accuracy and  
adequacy of RGB models by employing CMDs, spectroscopic observations,   
and luminosity functions.  
\S 9 reviews the use of the TRGB as standard candle, and  
conclusions follow in \S 10.  
   
\section{RGB stellar evolution}   
   
\lq{Canonical}\rq  stellar models are usually defined as    
models computed by solving the equations of stellar structure in   
the assumption of spherical symmetry (e.g. Kippenhahn \& Weigert~1991),    
neglecting magnetic fields, rotation and mass loss from the surface,    
considering convection as the only efficient mixing mechanism,    
and assuming that the convective regions are always fully mixed,     
their boundaries being fixed by the Schwarzschild criterion.   
In the rest of this paper we will use the definition
\lq{non-canonical}\rq  for models computed including additional 
physical effects like, e.g.   
overshooting from the Schwarzschild boundary of convective regions or   
atomic diffusion.    
   
The solution of the stellar structure equations for a given initial   
value of the total mass and chemical composition provides the run of   
physical quantities (such as density, radius, pressure, temperature,   
luminosity) as well as the chemical abundance profiles from the center up to the   
surface of the star, and their evolution with time.   
In order to compare theory with photometric observations the stellar   
evolution results must be then supplemented with some prescriptions to   
determine bolometric corrections and color indices, all along the   
evolutionary track.   
   
In the following we present a summary of the RGB stellar evolution (we refer the
reader to the seminal papers by Sweigart \& Gross~1978 and Sweigart, Greggio \&
Renzini~1989, 1990 for a detailed discussion on this subject).  
   
   
\subsection{Evolution up to the RGB tip}   
   
RGB stars are objects with masses lower than    
$\sim$2.0 $M_{\odot}$ (the precise value depends on the initial   
chemical composition), which develop electron   
degenerate He-cores after the end of central H-burning, surrounded by   
a thick H-burning shell (thickness of the order of 0.1$M_{\odot}$)   
and a convective envelope whose chemical composition   
is the initial one. The envelope temperature gradient is    
to a large extent adiabatic, apart from the most external layers,   
where it becomes superadiabatic, and must be treated according to some   
prescribed convection theory.   
   
At this stage, the convective    
envelope progressively deepens and the H-shell narrows down    
to a thickness of the order of 0.001   
$M_{\odot}$, while the star crosses the CMD towards redder colors   
(i.e. lower $T_{\rm eff}$ and larger radius).    
There is a large body of literature devoted to the identification of   
what is (or what are, if more than one) the precise physical reason    
for the expansion of the star to red giant dimensions (a fact that   
comes out naturally from the integration of the stellar structure   
equations), but a general   
consensus has not been yet reached; we address the reader to the    
papers by  Weiss~(1989), Iben~(1993), Renzini \&   
Ritossa~(1994), Sugimoto \& Fujimoto~(2000), and references therein.   
   
Due to the steady deepening of the convective region, the lower   
boundary of the convective envelope enters regions chemically   
processed (some produced He and the C and N abundances reaching their   
CN-cycle equilibrium value)   
by the central H-burning phase; this material is mixed almost instantaneously   
throughout the convective envelope, thus altering the   
surface abundances. This phenomenon, called first dredge-up,   
causes an increase of the surface He abundance, and an increase of the   
N/C ratio with respect to the original values.   
Convection reaches its maximum extension (in mass)    
near the base of the RGB\footnote{It is important to remark here   
that the very deep convective envelope is able to almost completely   
erase the effect of atomic diffusion --    
if efficient during the previous main sequence phase -- because it   
reengulfs almost all the material previously diffused out   
of the narrower convective region (and moreover the RGB evolution is   
too fast to be itself affected by atomic diffusion).    
This means that RGB models computed   
with and without considering atomic diffusion are almost   
indistinguishable (e.g., Proffitt \& Vandenberg~1991,   
Castellani et al.~1997). More in detail, the RGB $T_{\rm eff}$ 
is basically unaffected by the inclusion of diffusion, the 
$M_{core}^{He}$ values at the TRGB 
differ by about 0.004$M_{\odot}$ (larger values in the case of models with diffusion), 
and the TRGB surface He mass fraction is 
larger by $\sim$0.01 in the case of canonical models
(Cassisi et al.~1998).},   
and from this moment on the lower convective    
boundary slowly recedes towards   
the surface, due to the steady growth of the He-core which accretes He   
produced by the H-burning shell. The evolution along the RGB covers   
a relatively small range in $T_{\rm eff}$ ($T_{\rm eff}$ slowly   
decreasing with time) and a large range   
of surface luminosity, which increases with time.    
Due to the growth of $M_{core}^{He}$, the H-shell will   
encounter the sharp discontinuity of the H-profile left at the point of   
maximum extension of the convective envelope.   
The star reacts to the sudden increase of available fuel by   
lowering its surface luminosity and slowing down the evolutionary   
timescale, before starting again to increase its luminosity   
after the shell has moved past the discontinuity.   
This occurrence is recorded in both the differential luminosity
function (LF -- number of stars in a given brightness interval 
as a function of the brightness   
itself) and the integrated LF (sum of the number of stars from an   
arbitrary origin to a given value of the brightness, as a function   
of the brightness itself) of old stellar    
populations as, respectively,  a   
local peak (the so-called RGB bump) in the differential LF,   
or a break in the slope of the integrated LF.   
The RGB bump is  a genuine prediction of theory (Thomas~1967, Iben~1968b)   
confirmed many years later by observations of the globular cluster (GC) 47Tuc   
(King, Da Costa \& Demarque 1985).   
   
\begin{figure}   
\plotone{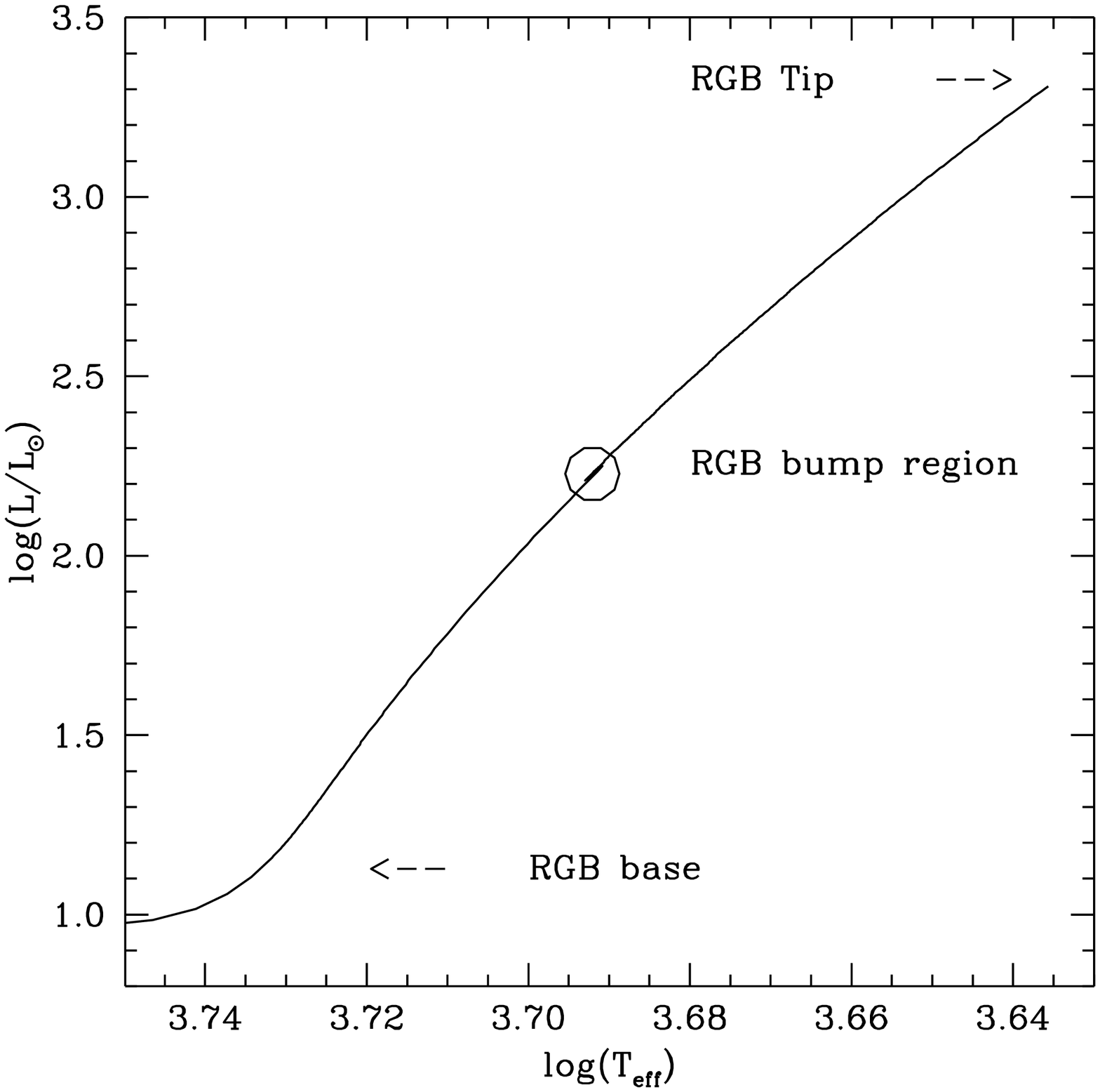}   
\caption{RGB evolution in the log($L/L_{\odot}$)/$T_{\rm eff}$  
plane of a 1$M_{\odot}$ star, Z=0.0004 and Y=0.231,   
scaled solar metal distribution.\label{hr}}   
\end{figure}

When $M_{core}^{He}$ reaches about 0.50 $M_{\odot}$ (the precise   
value depends weakly on the total mass of the star, being more sensitive    
to the initial chemical composition), He-ignition occurs in the   
electron degenerate core,   
producing the so called He-flash which terminates the RGB phase   
by removing the electron degeneracy in the core,    
and drives the star onto its Zero Age Horizontal Branch (ZAHB) location, that   
marks the start of quiescent central He-burning plus shell H-burning.   
As an example, Fig.~\ref{hr}  shows the evolution of a 1.0 $M_{\odot}$   
(Z=0.0004, Y=0.231) star from the base   
of the RGB up to the He-flash (Salaris \& Weiss~1998 models).    
The RGB bump region is marked with a   
circle; the RGB base and the TRGB are also indicated.   
   
Due to the relatively short timescale of RGB evolution   
(RGB timescales are of the order of a few percent    
of the main sequence lifetime), the RGB portion of theoretical isochrones    
(e.g. the CMD locus occupied by stars of   
different masses with the same initial chemical composition and age)    
is populated by stars all with almost the same mass. This means that   
for RGB stars in simple stellar populations (single age, single initial chemical   
composition), the CMD location and the expected number of stars along the 
red giant phase provided by an
isochrone of a fixed age, are equivalent to the those given by the    
RGB evolutionary track of the appropriate stellar mass.   
      
\subsection{Mass loss along the RGB}   
   
Stars along the RGB lose mass from their convective envelope (e.g.,   
Reimers~1975a,b), and the precise amount   
of mass loss along the RGB is a key parameter which determines   
color location and extension of the observed HB   
in GCs.   
{From} the point of view of the RGB evolution, however, mass loss    
(in the case of the rates necessary to explain the bulk of the observed    
HB stars in Galactic GCs, excluding the stars in the HB blue tails of   
some cluster, for which very extreme rates must be   
assumed -- see, e.g., Castellani \& Castellani~1993, D'Cruz et al.~1996)   
has a negligible effect on the overall properties of RGB stars.    
In fact, the stars react on Kelvin-Helmholtz timescales
to the surface mass loss, rearranging their radii  
according to the instantaneous value of the   
total mass (e.g. Castellani \& Castellani~1993); 
the stellar radius    
is however very weakly dependent on the value of the mass.   
The internal structure and evolutionary timescales   
are always completely unaffected by this process; this property is due to the 
fact that the nuclear timescale is significantly longer 
than the Kelvin-Helmholtz timescale of the envelope thermal readjustment. 
   
A typical value for the mass lost along the RGB in galactic GCs is of    
the order of 0.20$\rm M_{\odot}$ (as inferred from the observed color    
distribution of stars along the HB of the bulk of GCs);   
RGB tracks with this mass difference are almost coincident in the CMD.   
Moreover, the mass loss appears to be also a strong function of the stellar   
luminosity, so that the bulk of the RGB mass loss happens only close   
to the TRGB. A very informative summary of the available analytical mass loss formulae    
is presented in the appendix of Catelan~(2000).   
   
It is also worth mentioning that Soker et al.~(2001) have recently suggested   
the possibility that  stars just starting to move off the RGB toward the ZAHB location  
may undergo a super-wind phase, analogous to that observed in  
Asymptotic Giant Branch stars. This occurrence could provide a physical   
explanation for the gap in the HB stellar distribution  
at $T_{\rm eff}\sim 20000$ K, observed in several galactic GCs.

\subsection{The He-flash phase}   
   
As previously discussed, during the RGB evolution $M_{core}^{He}$ increases    
as a consequence of the He produced by the H-burning shell; since the core radius stays   
practically constant, the core density    
increases, and the resulting gravitational energy release   
raises the core temperature. In addition to this, another factor contributing to the   
increase of the core temperature is the steady increase with time of the temperature   
of the H-burning shell. The neutrino energy losses    
from the degenerate core also increase during the evolution. Although    
they do not prevent a continuous growth of the core temperature,    
when $M_{core}^{He}$ is larger than $\approx0.30M_\odot$, they are responsible   
for a temperature inversion in the core, so that    
the temperature maximum is no longer at the stellar center but at   
some distance from it.    
   
Due to the strong    
dependence of the 3$\alpha$ reaction rate on temperature,   
when the core gets hot enough, He-burning ignites off-center, close to    
the point where temperature is at its maximum.   
At the moment of the He-ignition in the core,   
density and temperature have attained values of the order of $10^6$ g $\rm cm^{-3}$    
and $8\cdot10^7$~K, respectively.    
In electron degeneracy conditions the gas pressure does not depend on   
temperature, so that an expansion does not   
immediately follow the local temperature increase caused by the   
He-burning energy release. As a consequence, at first  
the rate of burning, not being limited by a core expansion, becomes   
larger and larger and a thermal runaway ensues, the so-called He-Flash    
(see, e.g., Iben \& Renzini~1984 for a detailed discussion on the
subject). However, during the He-flash, the temperature increase causes   
the degree of degeneracy to decrease, so that     
a core expansion can begin. At the same time, as the energy flux in the   
core increases significantly, convection sets in, efficiently carrying out   
the produced energy. These two factors contribute to   
moderate the strength of the flash so that the star is not 
necessarily destroyed; the efficiency of the convective energy transport is especially
critical (see, e.g., Sugimoto~1964).
Recent hydrodynamical simulations by Achatz~(1995) and
Deupree~(1996) have indeed shown how
the flash does not produce a hydrodynamic event.   
The observational evidence provided by the   
existence of HB stars is the best proof   
that the overwhelming majority of stars undergoing He-flash do not explode.   
   
Another interesting question concerning the He-flash is the
possibility of partial or  
total mixing between the He core and the envelope. A long time ago    
it has been recognized (Schwarzschild \& H\"arm   
1962, 1964, 1967, Demarque \& Mengel 1972, Mengel \& Gross 1976,   
Renzini \& Fusi Pecci 1988 and references   
therein) that in low-mass stars    
such mixing does not take place, but different   
conclusions have been reached in the case of metal free (Pop~III) or   
extremely metal poor stars.   
Fujimoto, Iben \& Hollowell~(1990), Hollowell, Iben \& Fujimoto~(1990)   
and Fujimoto, Ikeda \& Iben~(2000)   
found that, due to the  small entropy barrier between the H- and the He-rich regions,   
the convective zone produced by the huge energy release of   
He-burning during the He-flash can penetrate the overlying H-rich layers.    
The resulting inward migration of protons into high-temperature   
regions leads to a H-shell flash, which further increases the extension   
of the central convective region.   
At the late stages of this phase the convective envelope   
deepens, and merges with the convective zone. Therefore, the initially metal free   
surface is enriched by a large amount of matter processed via He- and H-burning    
reactions, in particular a very high carbon abundance follows.   
A deeper insight on this phenomenon is very important, since it provides    
an attractive working hypothesis for explaining the peculiar chemical patterns observed in    
extremely metal-deficient stars.    
   
This topic has been recently addressed again by    
Schlattl et al.~(2001, see also Weiss et al.~2000a).   
They have investigated in detail the evolution during the He-flash   
phase of metal-free stars, covering a   
larger parameter space with respect to the works by Fujimoto and coworkers.    
Schlattl et al. (2001) have confirmed earlier results pointing out   
that the crucial parameter   
which defines whether the He-flash induced mixing occurs, is the location of the   
point where the He-burning energy release is maximum.    
The deeper inside the star this point lies, the lower is   
the probability that the convective zone developing at the He-flash   
ignition can penetrate into the hydrogen-rich    
envelope, thereby carrying down protons and triggering a H-burning runaway.   
   
However, even if the development of the process and its global   
properties are in fine agreement with the   
outcomes obtained by previous investigations, there is a fundamental difference:   
whereas Fujimoto et al.~(1990) and Fujimoto et al.~(2000) claim   
that the H-flash is a common property of all low-mass ($<1.0M_\odot$)   
metal-deficient stars, Schlattl et al.~(2001) have found that the   
occurrence of a  H-flash critically depends   
on the adopted initial conditions, like stellar mass and He abundance, as well   
as on the inclusion and/or assumptions made for the   
treatment of additional physical processes like atomic diffusion and   
external pollution of metals during the main sequence phase.    
   
A preliminary comparison with some extremely metal-poor stars, such as   
CS22892-052 and CS22957-027 (see Schlattl et al.~2001 for details),    
revealed a good match of effective temperature, luminosity and the   
relative abundances of  carbon and   
nitrogen, but too large absolute abundances of these elements.    
Even if the H-flash phenomenon provides an interesting working   
scenario for interpreting  the observed   
abundance patterns in extremely metal-deficient stars, it is not yet   
able to provide a detailed match to    
the observations. Whether this occurrence is a proof of the fact that the   
peculiar chemical abundances observed in   
extremely metal-poor stars may be primordial (hence these objects are   
not low mass Pop~III stars), due to external pollution at a late evolutionary   
stage, or a signature   
of drawbacks in current theoretical models for the   
He-flash phase in metal-free RGB stars, is not yet clear.   
\footnote{Schlattl et al.~(2001) have discussed some    
interpretations of this failure; one possibility is related   
to the crude assumptions made for deriving the velocity   
of the convective elements in the mixing region   
(we refer to the quoted paper for more details).}   
   
\section{Properties of the CMD of RGB stars}   
   
The behavior of RGB models in the observational CMD follows closely    
their properties in the theoretical luminosity-$T_{\rm eff}$ diagram.   
We have considered, as an example, the    
$VI$ (Johnson-Cousins) plane, and used the theoretical isochrones by    
Salaris \& Weiss~(1998) transformed into the $VI$ plane according    
to the Alonso, Arribas \& Martinez-Roger~(1999) empirical   
color transformations\footnote{Alonso et al.~(1999)    
provide $I$ magnitudes in the Johnson band. The transformation to the   
Cousins system has been performed following the prescriptions by Fernie~(1983).}.    
   
\begin{figure}   
\plotone{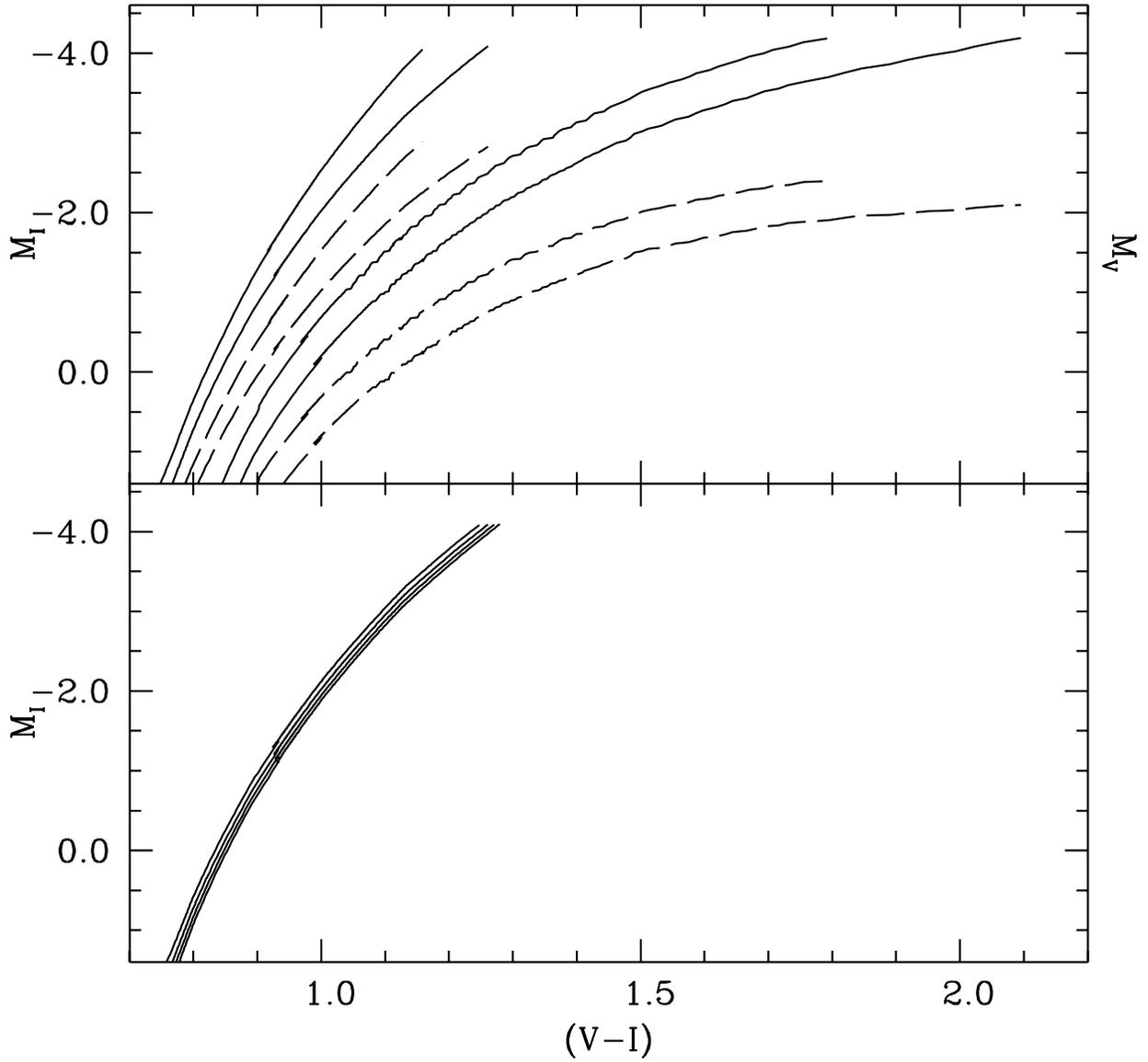}   
\caption{RGB isochrones in the $I-(V-I)$ Johnson-Cousins plane   
(solid lines) and in the $V-(V-I)$ plane (dashed lines). The   
upper panel shows 10 Gyr isochrones with   
Z=0.0002, 0.0004, 0.004, 0.008 (from left to right).    
The lower panel displays    
Z=0.0004 isochrones with ages of 8, 10, 12, 14 Gyr.\label{cmd1}}   
\end{figure}

The lower panel of Fig.~\ref{cmd1} shows the location in the $I-(V-I)$ plane of   
RGB isochrones of different ages and the same initial metal content,   
while the upper panel displays isochrones of the same age and   
different initial metallicities.    
It is evident that the RGB color at   
a given absolute brightness (hence the stellar $T_{\rm eff}$)   
is very weakly affected by the age of the parent population (hence   
by the value of the stellar mass), at least for ages higher than a few Gyr.    
On the other hand, it   
is strongly affected by the initial value of the metallicity Z (metals  
mass fraction). More metal rich   
RGB models are redder (see upper panel of Fig.~\ref{cmd1}).   
This property is, from the qualitative point of view,    
absolutely general and does not depend on the   
specific color bands used, since it mirrors the behavior of the   
models in the luminosity-$T_{\rm eff}$ plane.   
However, the precise value of the   
sensitivity of RGB colors to Z does depend on the wavelength   
bands employed in the CMD.    
   
Not only the color, but also the shape of the RGB is affected by Z,    
in a way which is    
strongly dependent on the wavelength bands used.   
This is clearly visible in the upper panel    
of Fig.~\ref{cmd1}, when the isochrones are plotted    
in the $V-(V-I)$ plane (dashed lines).   
In this plane the shape of the RGB is more dramatically   
affected, due to the behavior of the bolometric correction to the $V$ band    
as a function of metallicity.   
   
The shape and location of RGB models is   
almost unaffected by the selected He mass fraction Y, but it is    
sensitive to the metal distribution.   
In particular, as discussed by Salaris, Chieffi \& Straniero~(1993 --   
see also Salaris \& Cassisi~1996, Salaris \& Weiss~1998,    
Vandenberg et al.~2000),   
the abundance of low ionization potential elements like Mg, Si, S, Ca   
and Fe affects strongly the RGB temperature. In general, for typical   
Pop~II low mass stars, RGB models   
computed with an arbitrary heavy element mixture will be equivalent to   
models with scaled solar metal ratios of the same total metallicity,   
as long as the ratio $\frac{X_C+X_N+X_O+X_{Ne}}{X_{Mg}+X_{Si}+X_{S}+X_{Ca}+X_{Fe}}$   
(where $X_i$ is the mass fraction of the element $i$)    
in the selected metal mixture is the same as in the scaled solar one.   
Since what is measured spectroscopically is [Fe/H] and [$\alpha$/Fe],   
it is useful to consider the following relationship (Salaris et   
al.~1993) linking an input parameter of theoretical models like   
[M/H] ([M/H]$\sim$log(Z/$\rm Z_{\odot}$)) 
to [Fe/H] and $f_{\alpha}=10^{[\alpha/Fe]}$,   
which holds in case of the scaled solar metal   
mixture by Ross-Aller~(1976):   
   
\begin{equation}   
\rm [M/H]\sim \rm [Fe/H]+\rm log(0.638 \sl f_{\alpha}+ \rm 0.362)   
\end{equation}   
   
Small adjustments to the   
coefficients must be applied in case of more recent    
determinations of the solar metal distributions; Yi et al.~(2001) have  
obtained, in case of the Grevesse \& Noels~(1993) solar metal  
distribution:  
  
\begin{equation}   
\rm [M/H]\sim \rm [Fe/H]+\rm log(0.694 \sl f_{\alpha}+ \rm 0.306)   
\end{equation}

As far as the RGB location is concerned, this equivalence   
between scaled solar and $\alpha$-enhanced models with the same total metallicity   
breaks down when Z is greater than $\sim$0.002 ([M/H]$\sim-$1).   
In case of the TRGB and RGB bump luminosities, and the subsequent   
ZAHB luminosity, the previously discussed agreement between scaled   
solar and $\alpha$-enhanced models with the same total metallicity   
holds for Z up to $\sim$0.01.     
   
As for the TRGB brightness, this is weakly dependent on the    
stellar mass (Fig.~\ref{cmd1}), and therefore on the isochrone age, for
ages larger than $\sim$4-5 Gyr.  
This is due to the fact that, at a given initial   
chemical composition, the TRGB level (and in general the stellar   
surface luminosity all along the RGB) is determined by the value of   
$M_{core}^{He}$, and its value at the He-flash is fairly  
constant over large part of the low-mass star range.   
$M_{core}^{He}$ decreases for increasing   
metallicity, while the TRGB brightness increases due to the increased   
efficiency of the H-shell, which compensates for the reduced core mass.   
The brightness of the subsequent ZAHB phase   
follows the behaviour of $M_{core}^{He}$, decreasing for increasing    
metallicity.   
   
$\rm M_{I}^{TRGB}$ appears to be also very weakly sensitive to the   
heavy element abundance   
(e.g., LFM93, Salaris \& Cassisi~1997); for [M/H] ranging between    
$-$2.0 and $-$0.6, $\rm M_{I}^{TRGB}$    
changes by less than 0.1 mag. In fact,    
$\rm M_{bol}^{TRGB}$ is proportional\footnote{We are considering   
[$\alpha$/Fe] ratios approximately  
constant with respect to [Fe/H].}    
to $\sim -$0.18[M/H], while $\rm BC_{I}$ is   
proportional to $\sim -0.24(V-I)$ (e.g., Da Costa \&   
Armandroff~1990). Since the color of the TRGB goes   
approximately as 0.57[M/H], then $\rm BC_{I}$ is proportional    
to $\sim -$0.14[M/H].    
The slope of the $\rm BC_{I}-$[M/H] relationship is therefore    
very similar to the slope of    
the $\rm M_{bol}^{TRGB}-$[M/H] relationship, and since    
$\rm M_{I}^{TRGB}$=$\rm M_{bol}^{TRGB}- \rm BC_{I}$, $\rm M_{I}^{TRGB}$   
is almost independent of the stellar metal content.   
This is the basis for its use    
as distance indicator. On the other hand, due again to the behaviour of the   
bolometric corrections to the $V$ band, $\rm M_{V}^{TRGB}$   
is strongly affected by the metal content of the stars.   
   
\section{Properties of the LF of RGB stars}   
   
We have already    
defined the differential LF (or, simply, the LF)   
of RGB tracks (or isochrones) as the run of the number   
of stars per brightness bin along the RGB, as a function of the   
brightness itself. The number of objects in a given bin is of course   
determined by the local evolutionary speed, so that the comparison   
of observed and theoretical RGB LFs provides a key test for the   
accuracy of the predicted RGB timescales.   
   
\begin{figure}   
\plotone{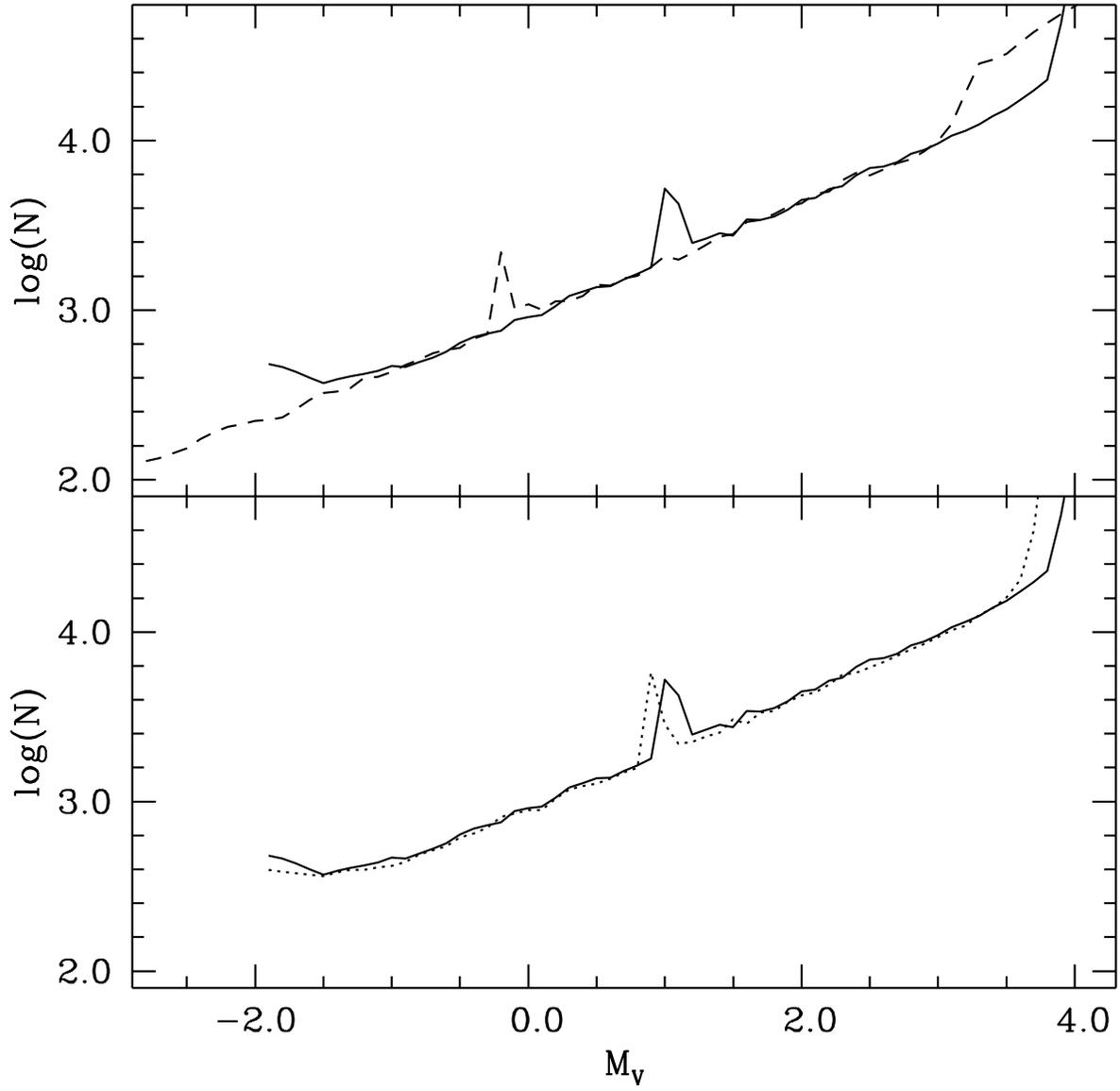}   
\caption{The upper panel shows a comparison of 13 Gyr LFs with Z=0.008   
(solid line) and Z=0.0004 (dashed line); the lower panel displays    
two LFs with Z=0.008 and t=10 (dotted line) and 13 (solid line) Gyr.\label{lf1}}   
\end{figure}

In Fig.~\ref{lf1} (upper panel) two V band RGB LFs    
(bin size of 0.1 mag), for isochrones of 13 Gyr and Z=0.0004   
(dashed line) and Z=0.008   
(solid line) are shown; they are normalized to the same    
number of stars between $\rm M_{V}$=2.0 and 3.0. The local maxima   
located at $\rm M_{V}\sim-$0.2 and 1.0 correspond to    
the RGB bump; the increase of the number of stars close to the TRGB    
of the more metal rich LF is due to the fact that the V magnitude of the   
RGB models at this metallicity tends to stay constant or increase   
slightly toward the TRGB (see Fig.~\ref{cmd1}).  
The overall linearity of the LF is a straightforward consequence of the existence 
of a luminosity - $M_{core}^{He}$ relation (see Eggleton 1968 and 
Castellani, Chieffi \& Norci 1989 for a detailed 
discussion about this point).
It is evident that the ratio of the evolutionary timescales along   
the RGB does not depend on Z, since the slope of the LFs
(i.e. the luminosity - $M_{core}^{He}$ relation) are    
almost coincident.    
   
In the lower panel of the same figure the effect of age (value of the   
evolving mass) is shown for two LFs with Z=0.008 and ages of,   
respectively, 10 (dotted line) and 13 (solid line) Gyr.   
The LF (like the CMD location)   
is basically unaffected by the age and    
only the position of the bump is slightly changed.   
We address the reader to the paper by Vandenberg, Larson \& de Propris~(1998) for
a more detailed discussion about the dependence of the LF on various physical
and chemical inputs.
   
\section{RGB input physics}   
   
The quality of the agreement between observations and theory is   
affected by both the adequacy of the assumptions built into   
the stellar models, and the accuracy of    
the description of the physical processes accounted for.   
In the following, we discuss in some detail the input physics used   
in the current generations of RGB models, highlighting the influence on   
RGB stellar properties of the various choices made by different authors.   
   
   
\subsection{Radiative opacity}   
   
Whenever the energy is transported by radiative processes a knowledge of   
the radiative opacity of the stellar matter is required in order to determine   
the temperature gradient. Within the star, where the diffusion   
approximation is applied to the energy transfer equation, the Rosseland mean   
opacity is required.   
We define here as low-temperature (low-$T$) opacities,    
Rosseland mean values when $T \leq 12000$ K, and high-temperature (high-$T$)    
opacities the values at larger temperatures. As discussed by Salaris  
et al~(1993), it is the low-$T$ opacities which mainly determine the   
$T_{\rm eff}$ location of theoretical RGB models, while the    
high-$T$ ones enter in the determination of the    
mass extension of the convective envelope (through the value of the  
radiative temperature gradient).   
   
Current generations of stellar models employ mainly   
the low-$T$ opacity calculations by Alexander \& Ferguson~(1994) --    
and in some cases the Kurucz~(1992) ones -- which constitute the most    
up-to-date computations suitable for stellar modelling,   
spanning a large range of initial chemical compositions.     
The main difference between these two sets   
of data is the treatment of molecular absorption, most notably the   
fact that Alexander \& Ferguson~(1994) include the effect of the $\rm H_{2}O$   
molecule. Salaris \& Cassisi~(1996) have compared, at different   
initial metallicities, stellar models   
produced with these two sets of opacities (as well as with the    
less used Neuforge~1993 ones, which provide results almost   
undistinguishable from models computed with Kurucz~1992    
data), showing that a very good agreement exists   
when $T_{\rm eff}$ is larger than $\sim$4000 K.    
As soon as the RGB $T_{\rm eff}$ goes below this limit    
(when the models approach the TRGB   
and/or their initial metallicity is increased),    
Alexander \& Ferguson~(1994) opacities produce progressively    
cooler models (differences reaching values of the order of 100 K or more),    
due to the effect of the $\rm H_{2}O$   
molecule which contributes substantially to the opacity in this   
temperature range.   
   
As for the high-$T$ radiative opacities, the OPAL results    
(e.g. Iglesias \& Rogers~1996) are generally    
used; these opacities, coupled with the OPAL equation   
of state (Rogers, Swenson \& Iglesias~1996), have been repeatedly   
tested against helioseismological constraints (e.g., Degl'Innocenti et al.~1997),
providing a good agreement between theory and observations, at least in the 
solar regime.   
   
\subsection{Electron conduction opacity}   
   
In the electron degenerate He-core an evaluation of the electron   
conduction (or, briefly, conductive) opacity is needed in order to   
determine the local temperature gradient. The precise computation    
of the conductive opacities is fundamental for deriving the   
correct value of the He-core mass at the He-flash,    
since, for example, higher conductive opacities cause a    
less efficient cooling of the   
He-core and an earlier He-ignition (i.e., at a lower core mass).   
   
Two main choices are presently available, neither of   
which is totally satisfactory:   
the analytical fitting formulae by Itoh et al.~(1983 -- I83),    
or the old Hubbard \& Lampe (1969 -- HL69) tabulation.   
As pointed out by Catelan, de Freitas Pacheco \& Horvath~(1996),   
the most recent results by I83 are an improvement over the older HL69   
ones, but their range of validity does not cover the He-cores of RGB stars.   
The I83 opacities are   
generally smaller than those obtained from the HL69 tables, thus implying   
larger values of $M_{core}^{He}$ at the He-flash. Castellani \&   
Degl'Innocenti~(1999) find an increase by 0.005$M_{\odot}$ of   
$M_{core}^{He}$ core at the He-flash for a 0.8$M_{\odot}$ model with initial   
metallicity Z=0.0002, while   
in case of a 1.5$M_{\odot}$ star with solar chemical composition the   
increase amounts to 0.008$M_{\odot}$ (Castellani et al.~2000).   
$M_{core}^{He}$ variations of this kind   
imply an increase of the predicted ZAHB   
brightness at the RR Lyrae temperatures in the range    
0.05-0.07 mag, and a slightly larger increase of the predicted TRGB brightness by   
$\sim 0.09$ mag.   
   
It is not obvious which is the most appropriate set of opacities to use,    
and an extension  of the I83 work to physical conditions typical    
of RGB cores is very much needed.   
   
\subsection{Equation of state}   
   
The equation of state (EOS) of the stellar matter     
is another key input for the model computations; it connects    
pressure, density, temperature and chemical composition at each point   
within the star, determines the value of the adiabatic   
gradient (which is the temperature gradient in most of the   
convective envelope), the value of the specific heat (which appears in the   
expression of the gravitational energy term),   
and plays a crucial role in the evaluation of    
the extension of the convective regions. 
   
The best available EOS is probably the OPAL one (the EOS used to   
compute the OPAL radiative opacities), which, as already mentioned,   
produces remarkably good results in the solar regime. However, its   
range of validity does not cover either the electron degenerate cores   
of RGB stars, or their cooler, most external layers, below 5000 K.   
RGB models computed with the OPAL EOS must employ some   
other EOS to cover the most external and internal stellar regions.   
As for the models discussed later,  
Caloi, D'Antona \& Mazzitelli~(1997) and Silvestri et al.~(1998) use    
the MHD (D\"appen et al.~1988) EOS below   
5000 K, while the Magni \& Mazzitelli~(1979) one is used for the cores;    
Salaris \& Weiss~(1998 -- SW98) and Cassisi et al.~(1998)  
complement the OPAL EOS with the Straniero~(1988) one for the   
degenerate cores and a Saha EOS for the most external layers,   
while the   
Yale-Yonsei models (Yi et al.~2001 -- YY01)   
employ the Yale EOS (e.g., Chaboyer \& Kim~1995) for   
regions outside the range of validity of the OPAL data.  
The widely used Padova models by Girardi et al.~(2000 - P00) employ   
their own implementation of the Straniero (1988) EOS for T$>10^7$ K 
(Girardi et al 1996), and the MHD EOS at lower temperatures. 
Vandenberg et al.~(2000 -- V00) models make use of the EOS described  
in the appendix of their paper, while Salaris \& Cassisi~(1998 -- SC98) RGB tracks   
use the EOS by Straniero~(1988) throughout all the stellar structure,   
plus a Saha EOS below $10^6$ K.   
   
No detailed study exists highlighting the effect of the various   
EOS choices on the evolution and properties of RGB stars.    
However, the EOS does affect the $T_{\rm eff}$ of stellar models in   
general and RGB ones in particular. After computing two 0.8$\rm   
M_{\odot}$ (Z=0.001) RGB tracks      
employing in the envelope either the OPAL EOS or the Straniero~(1988) plus   
Saha one at temperatures less than $10^{6}$ K (and   
keeping everything else fixed),   
we obtained differences of the order of more than 100 K, at least when   
$T_{\rm eff}$ is larger than about 5000 K and the envelope 
can be fully covered by the OPAL EOS.   
As far as the properties of TRGB models are concerned, a preliminary investigation on the effect
of different EOS choices has been performed by Vandenberg \& Irwin~(1997). They have shown 
that neglecting the Coulomb interaction and electron exchange increases 
$M_{core}^{He}$ at the He-flash by $\sim$0.005$M_{\odot}$ (see also
Harpaz \& Kovetz~1988).

   
\subsection{Nuclear reaction rates}   
   
The He-flash in the core of RGB stars occurs when the 3$\alpha$    
process is ignited;   
the rate of the 3$\alpha$ reactions is therefore important for   
determining the corresponding value of $M_{core}^{He}$. The    
rate by Caughlan \& Fowler~(1988) is widely employed; its estimated error is   
of about $\pm$15\% (see, e.g., the discussion in    
Castellani \& Degl'Innocenti~1999), which causes a negligible   
uncertainty on the value of $M_{core}^{He}$ at the He-flash, of about   
only $\pm$0.001$M_{\odot}$.    
If the previous rate by Fowler et al.~(1975) is used,   
$M_{core}^{He}$ is reduced by about 0.004$M_{\odot}$   
(Cassisi et al.~1998).   
   
Closely connected to the nuclear reaction rates is the evaluation of
the effect of the plasma screening in stellar conditions. The
screening factors usually   
employed are from De Witt, Graboske \& Cooper~(1973) and    
Graboske et al.~(1973). The appropriate regime for the He ignition   
in low mass stars is   
the intermediate one (e.g. Graboske et al.~1973).   
Using the formulas for the weak screening in place of the more appropriate   
intermediate one causes a reduction of $M_{core}^{He}$ at the flash   
by $\sim$0.006$M_{\odot}$ (Caloi, D'Antona \& Mazzitelli~1997).   
More recent developments do not modify   
appreciably the He-core masses obtained with this formalism   
(e.g. Catelan et al.~1996, Cassisi et al.~1998).   
   
\subsection{Neutrino energy losses}   
   
A precise determination of the energy losses due to neutrino emission   
is also important in order to determine precisely the core mass at the   
flash. Since neutrinos subtract energy from the stellar interior, they   
contribute to the cooling of the core and  
affect the value of core mass at He ignition (see, e.g., Sweigart \& Gross~1978).   
   
Plasma processes are the dominant   
mechanism for neutrino emission in the core of RGB stars; the best   
determination of their rate in RGB conditions is provided by   
Haft, Raffelt \& Weiss~(1994). Their formula improves the previous   
results by Munakata, Kohyama \& Itoh~(1985),   
which were widely used in stellar evolution calculations.   
Switching from the Munakata et al.~(1985) results to the Haft et   
al.~(1994) ones, produces an increase of $M_{core}^{He}$ at the He-flash   
of $\sim$0.005 - 0.006$M_{\odot}$ (Haft et al.~1994, Catelan et al.~1996,   
Cassisi et al.~1998).   
The quoted uncertainty on the Haft et al.~(1994) result is $\pm$5\%,   
corresponding to a negligible uncertainty on the core mass at the    
flash (Castellani \& Degl'Innocenti~1999).   
   
A comprehensive compilation of the most recent results on the various   
mechanisms of neutrino emission, suitable for use in stellar   
evolution computations (including the Haft et al.~1994 results),    
can be found in Itoh et al.~(1996).

\subsection{Treatment of surface boundary conditions}   
   
In order to integrate the stellar structure equations it is necessary   
to fix the value of the pressure and temperature   
at the surface of the star, usually close   
to the photosphere. There are basically two possibilities to do this:   
   
i) integrating the atmospheric layers by using a T$(\tau)$   
relationship, supplemented by the hydrostatic equilibrium condition   
and the equation of state;   
   
ii) obtaining the desired boundary conditions from pre-computed non-gray   
model atmospheres.   
  
\begin{figure}   
\plotone{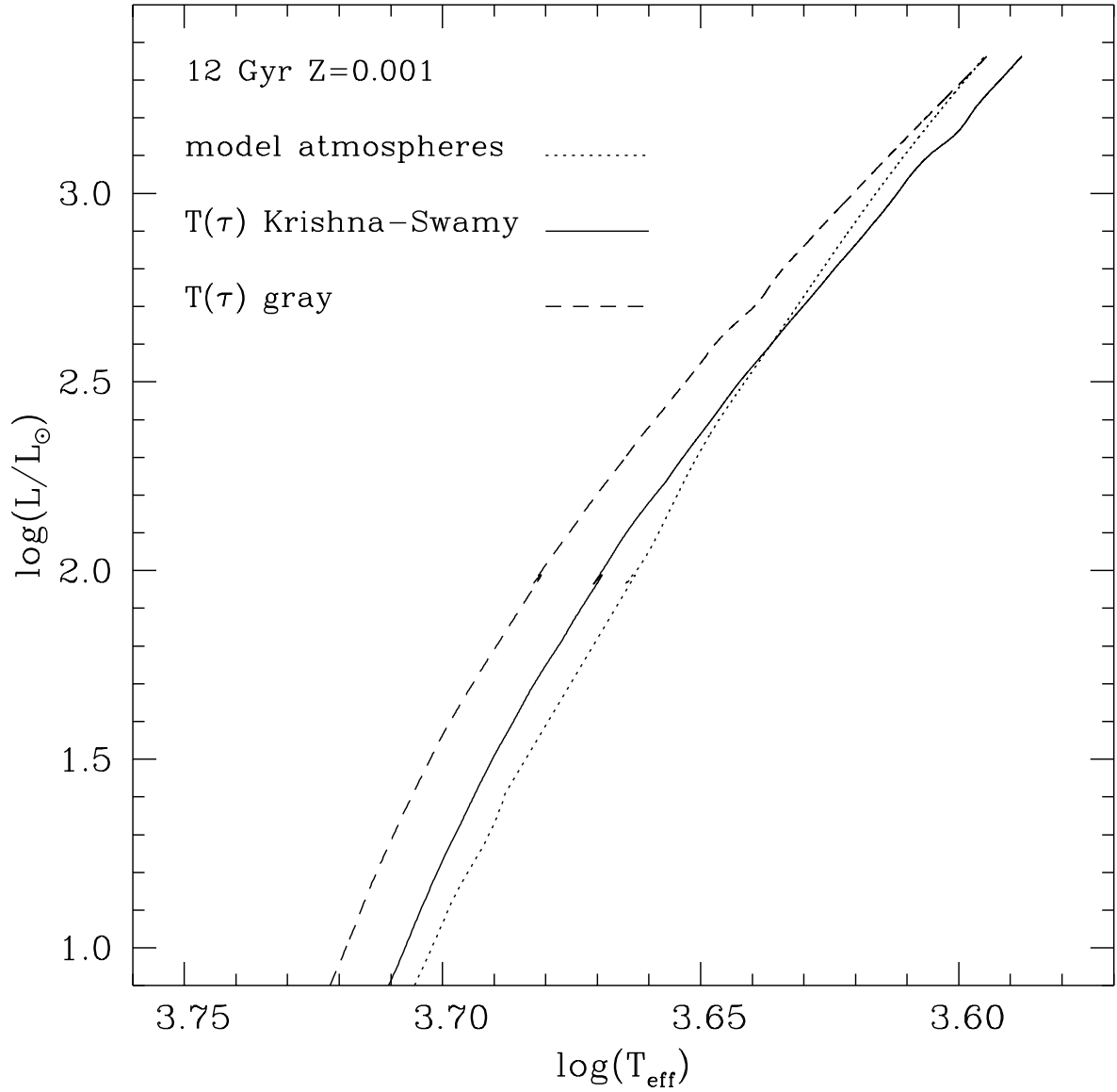}   
\caption{Comparison of 12 Gyr RGB models for Z=0.001, computed using    
model atmospheres (dotted line) and, respectively, gray (dashed line)  
and Krishna-Swamy~(1966 -- solid line) T($\tau$) relationships   
to obtain the surface boundary conditions. See text for details.   
\label{modatm}}   
\end{figure}

The first procedure is universally used in the current generation of   
stellar models. The effects on RGB stellar models of different $T(\tau)$ relations  
is shown in Fig.~\ref{modatm}, where the RGB of two 12 Gyr isochrones  
with Z=0.001 computed using, respectively, the Krishna-Swamy~(1966) solar T$(\tau)$  
relationship, and the gray one (see, e.g., Table 3.2 in Mihalas~1978), 
are displayed (all other parameters being kept fixed).  
RGBs computed with a gray  T$(\tau)$ are systematically hotter by  
$\sim$100 K.   
  
Recent studies of the effect of using boundary   
conditions from model atmospheres are in V00 and   
Montalban et al.~(2001). In the same Fig.~\ref{modatm}, as an example,    
we show also a RGB computed using  
boundary conditions from the Kurucz~(1992) model   
atmospheres, taken at $\tau$=10.    
The three displayed RGBs, for consistency,   
have been computed by employing the same  
low-T opacities, namely the   
ones provided by Kurucz~(1992), in order to be homogeneous with the model atmospheres.   
The model atmosphere RGB  
shows a slightly different slope, crossing over the   
evolutionary track of the models computed with the 
Krishna-Swamy~(1966) solar T$(\tau)$,   
but the difference with respect to the latter stays always within   
$\sim \pm$50 K.    
It is important to notice that the convection treatment in the   
adopted model atmospheres is not the same as in the underlying   
stellar models (i.e., a different mixing length formalism and a
different value for the scale height of the convective motion is used; see
next subsection for more details on the convection treatment). Montalban et al.~(2001)   
have discussed in detail the effect of boundary conditions from model   
atmospheres where convection is treated exactly as in the stellar   
models. Their tracks extend up to the lower part of the RGB, where   
their results are qualitatively similar to ours.

\subsection{Treatment of convection}   
   
The temperature gradient along the bulk of the convective envelope    
can be reliably approximated by the value of the adiabatic gradient;   
however, in the layers close to the stellar surface the convective   
gradient becomes strongly superadiabatic. There is no   
doubt that the determination of the temperature gradient in these   
regions is one of   
the most important unsolved problems of stellar evolution.   
The mixing length theory (MLT; B\"ohm-Vitense 1958) is    
almost universally used.    
It contains a number of free parameters, whose   
numerical values affect the model $T_{\rm eff}$; one of them  
is $\alpha_{\rm MLT}$, the ratio of the mixing length to the pressure scale   
height, which provides the scale length of the convective motions  
(increasing $\alpha_{\rm MLT}$ increases the model $T_{\rm eff}$).   
There exist different versions of the MLT, each one assuming different   
values for these parameters; however, as demonstrated by Pedersen,   
Vandenberg \& Irwin~(1990), the $T_{\rm eff}$ values obtained from the   
different formalisms can be made consistent, provided that a suitable value of   
$\alpha_{\rm MLT}$ is selected. Therefore, at least for the evaluation of    
$T_{\rm eff}$, the MLT is basically a one-parameter theory.   
   
The value of $\alpha_{\rm MLT}$   
is usually calibrated by reproducing the solar $T_{\rm eff}$, and this   
solar-calibrated value\footnote{We notice that the solar-calibrated $\alpha_{\rm MLT}$ increases 
by about 0.1 if element diffusion is accounted for. In the following we always refer to a 
$\alpha_{\rm MLT}$ solar calibration without including diffusion.} is then used for computing models 
of stars very   
different from the Sun (e.g. metal poor RGB and main sequence stars of   
various masses). It is clear that, by using the MLT,   
even if the input physics employed  
in the model computation is not accurate, it is possible to mask this  
shortcoming -- at least from the point of view of the predicted  
$T_{\rm eff}$ -- by simply recalibrating $\alpha_{\rm MLT}$ on the Sun. This  
guarantees that the models always predict correctly the $T_{\rm eff}$  
of at least solar type stars. However, since    
the RGB location is much more sensitive to the value of $\alpha_{\rm MLT}$   
than the main sequence,  
it is important to     
be sure that a solar\footnote{The solar calibrated   
numerical value of $\alpha_{\rm MLT}$ depends not only on the adopted input physics,   
but also, as previously discussed, on the selected MLT formalism.   
For these reasons, to allow a comparison of the results by   
different authors, the adopted MLT formalism has to be specified   
in addition to the input physics and     
boundary conditions employed.} $\alpha_{\rm MLT}$ is always suitable also for RGB stars of   
various metallicities.   
   
Detailed 2-dimension radiation hydrodynamics simulations of   
stellar envelope convection spanning a large range of gravities,     
$T_{\rm eff}$ and initial chemical compositions,   
have been recently carried out by Ludwig,   
Freytag \& Steffen~(1999), and applied to the case of metal poor   
low-mass stars by Freytag \& Salaris~(1999). These simulations   
(which cover a portion of the main sequence and the lower part    
of the RGB), within the limit of    
their employed input physics and numerical accuracy,   
show that the $T_{\rm eff}$ of the Sun and RGB stars of various    
metallicities   
can be well approximated by the MLT with a constant value of $\alpha_{\rm MLT}$.   
   
\begin{figure}   
\plotone{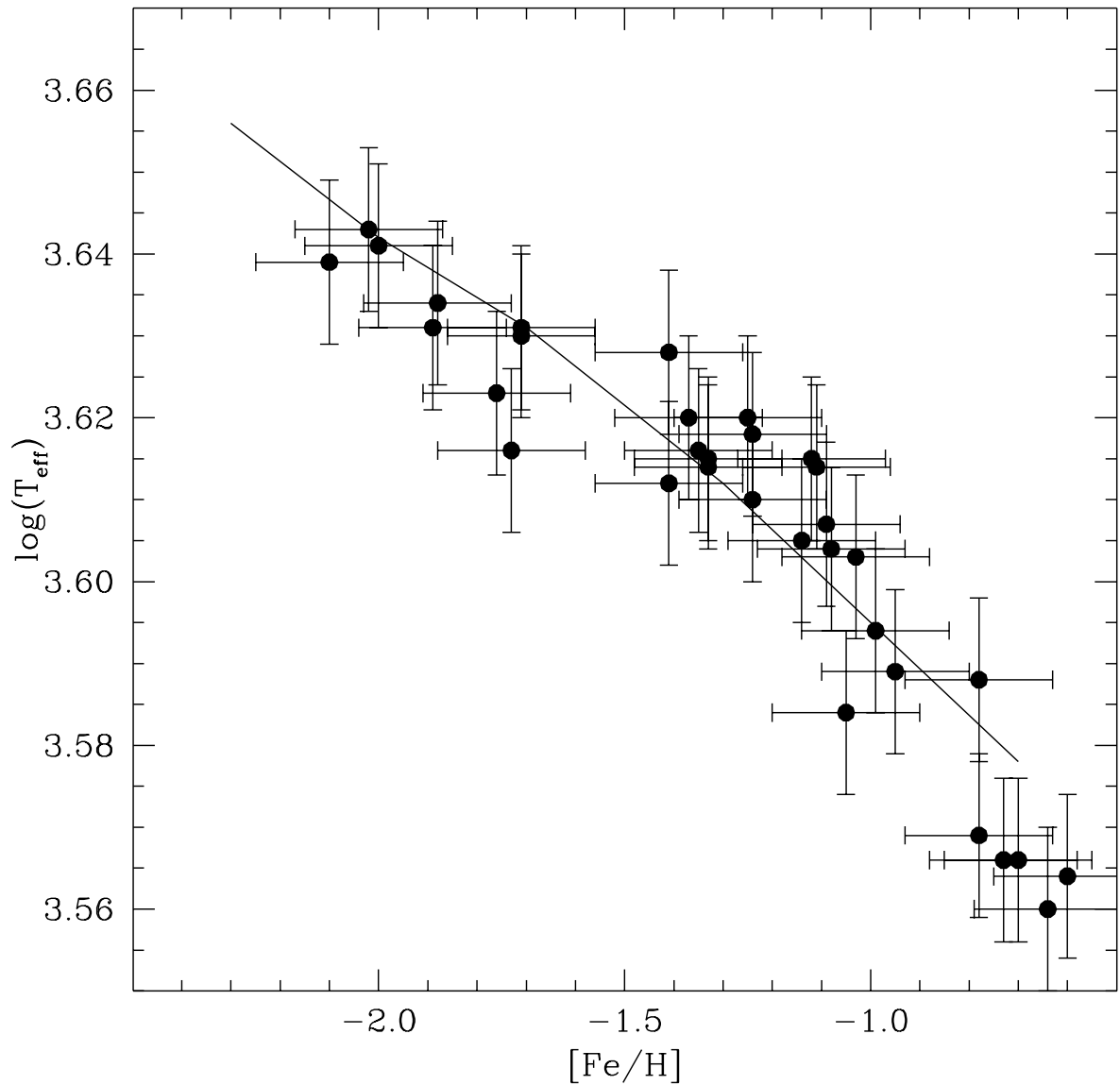}   
\caption{RGB $T_{\rm eff}$ (at $\rm M_{bol}=-$3) as a function of [Fe/H] from    
Salaris \& Weiss~(1998) 12 Gyr $\alpha$-enhanced ([$\alpha$/Fe]=0.4)     
isochrones (solid line), compared with globular clusters $T_{\rm eff}$    
from Frogel, Persson \& Cohen~(1983), and [Fe/H] from Carretta \&   
Gratton~(1997). Displayed error bars are of 100 K in $T_{\rm eff}$   
and 0.15 dex in [Fe/H].\label{calib}}   
\end{figure}

An independent way of calibrating $\alpha_{\rm MLT}$ for RGB stars is to compare   
empirically determined RGB $T_{\rm eff}$ values for galactic GCs   
with theoretical models of the appropriate chemical   
composition (see also Salaris \& Cassisi~1996,   
Vandenberg, Stetson \& Bolte~1996 and references therein).   
In Fig.~\ref{calib}, as an example, we show a comparison between  the  
$T_{\rm eff}$ from Frogel, Persson \& Cohen~(1983 -- FPC83)  
for a sample of GCs and the $\alpha$ enhanced models by SW98.  
In principle, the distance scale plays a role,    
since one has to know the clusters' distances in order to compare observed and   
computed RGBs; however, since the RGB is roughly vertical,   
uncertainties of the order of 0.2 mag in the GC distance moduli    
do not substantially influence the calibration.    
Just to give a measure of the sensitivity of the calibrated $\alpha_{\rm MLT}$   
value to temperature scale, chemical composition and adopted   
distances, we notice that systematic   
changes of the empirical $T_{\rm eff}$ by $\sim$70 K, or the cluster   
[M/H] by 0.2 dex, or the adopted cluster distance moduli by 0.25   
mag, cause a variation of the calibrated $\alpha_{\rm MLT}$ by $\sim$0.1.   
The GC reddenings have to be known to determine   
$T_{\rm eff}$ from the cluster photometry;   
when estimating the empirical $T_{\rm eff}$ with the Infrared Flux   
Method (IRFM), however, an uncertainty of $\pm$0.02 mag in the   
adopted GC reddening causes a variation of the calibrated $\alpha_{\rm MLT}$ by less   
than 0.05 (e.g. Alonso et al.~2000).   
   
Very recently Alonso et al.~(2000), using the theoretical $\alpha$-enhanced   
models by SW98, empirical estimates of $T_{\rm eff}$    
based on the IRFM (Alonso, Arribas \& Martinez Roger~1999),   
and the Carretta \& Gratton~(1997)   
[Fe/H] scale (together with a constant [$\alpha$/Fe]=0.4 built into the   
theoretical models), concluded that the IRFM   
temperature of RGB stars in clusters spanning a wide metallicity range   
is adequately reproduced by theoretical models computed by adopting   
a mixing length value within $\pm$0.1 of the solar one.   
Also V00 found a good agreement between their solar $\alpha_{\rm MLT}$, [$\alpha$/Fe]=0.3  
models and RGB $T_{\rm eff}$ from FPC83,  
for the GCs M92, M13 and 47Tuc ([Fe/H] between $\sim-$2.3 and $\sim -$0.7).  
It is worth noting that the FPC83 temperature scale differs from   
the Alonso et al.~(2000) one by only $\sim$50 K.  
  
These results seem to point to the fact that,    
as a general rule, the solar $\alpha_{\rm MLT}$ value is {\sl a priori}  
adequate also for RGB stars.    
Caution is however still necessary,   
for the following three main reasons. Regarding the   
hydrodynamical-simulations, the models reviewed before are not yet
full 3D simulations,   
extend only up to the RGB base, and, as discussed e.g. in   
Zahn~(1999), they are still unable to   
resolve all scales of convective motions.   
As for the calibration using empirical temperatures of RGB stars,    
one has to deal with the fact that it rests on the accuracy of    
both the $T_{\rm eff}$ and [Fe/H] scale, together with the   
determination of the metal distribution. Systematic uncertainties    
on the GC [Fe/H] scale are probably on the order of 0.2 dex    
(see, e.g., Rutledge, Hesser \& Stetson~1997), while     
uncertainties on the empirical $T_{\rm eff}$ scale are harder to   
evaluate. If color-$T_{\rm eff}$ relationships from theoretical model atmospheres   
are used to derive stellar temperatures, one has to face the   
associated uncertainties of the order of hundreds of Kelvin   
(see, e.g., Weiss \& Salaris~1999). It is however important to notice  
that, as previously mentioned, empirical determinations   
based on both the IRFM (Alonso et al.~1999), or on    
empirical $(V-K)$-$T_{\rm eff}$ relationships (FPC83),   
differ on average by only about 50 K.   
   
\begin{figure}   
\plotone{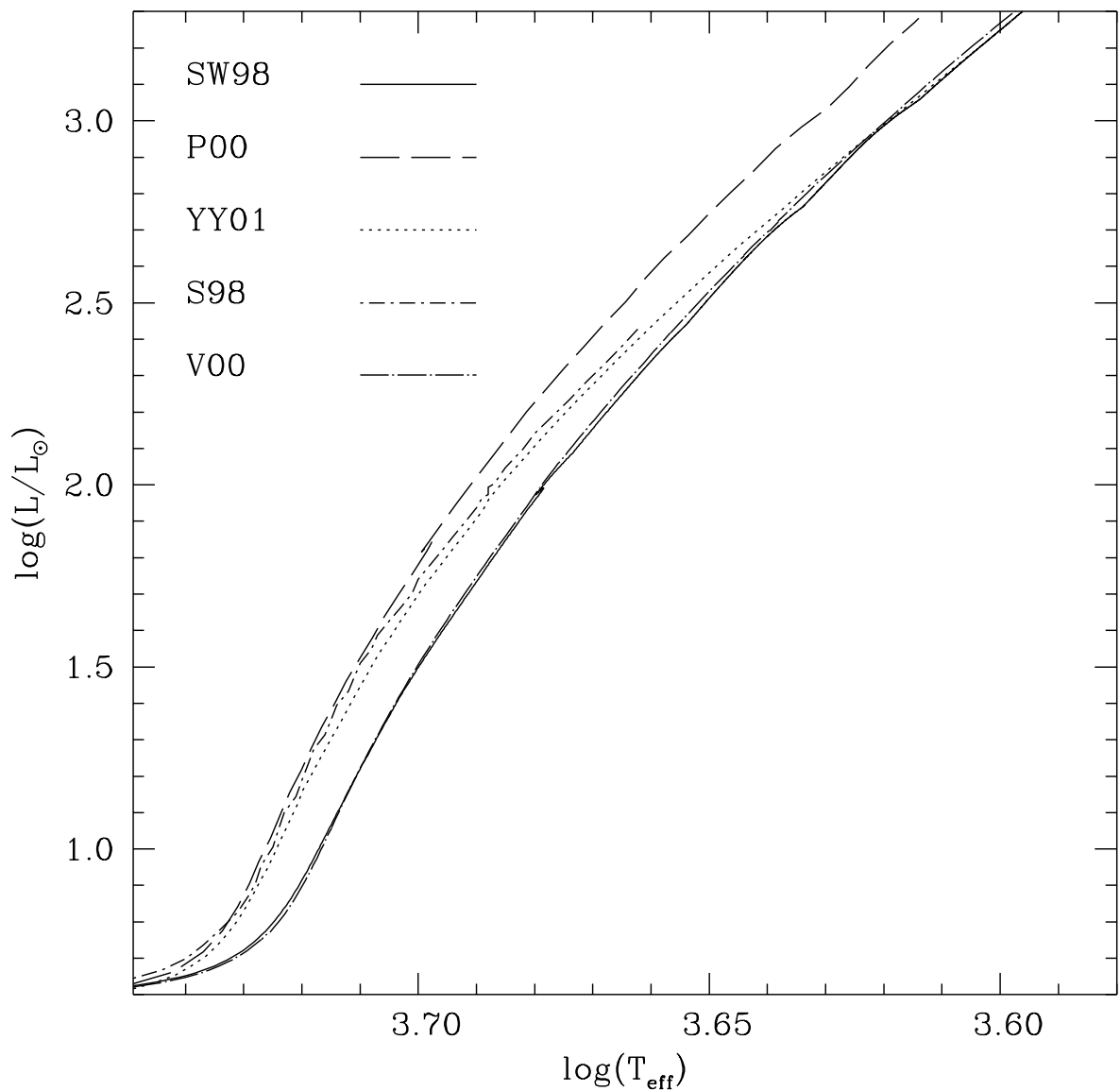}   
\caption{RGB isochrones computed with a solar calibrated $\alpha_{\rm MLT}$  
for a scaled-solar Z=0.001 metal mixture   
and t=12 Gyr, from different sources: Girardi et al.~(2000 -- P00); Yonsei-Yale   
models (Yi al.~2001 -- YY01); Vandenberg et al.~(2000 -- V00);  
scaled solar models computed with   
the same input physics as in Salaris \& Weiss~(1998 -- SW98); FST   
models by Silvestri et al.~(1998 -- S98).   
\label{rgbcomp}}   
\end{figure}

Another source of concern about an {\sl a priori} assumption of a solar $\alpha_{\rm MLT}$   
for RGB computations comes from the fact that   
recent models from various authors, all using a suitably     
calibrated solar value of $\alpha_{\rm MLT}$, do not show the same RGB   
temperatures.    
This means that -- for a fixed RGB temperature scale --   
the calibration of $\alpha_{\rm MLT}$ on the empirical $T_{\rm eff}$ values   
would not provide always the solar value.   
Figure~\ref{rgbcomp} displays four isochrones produced by different  
groups (SW98, P00, V00, YY01; also the Silvestri et al.~1998 models are
shown, but they will be discussed later), all   
computed with the same initial chemical composition, same opacities,    
and the appropriate solar calibrated values of   
$\alpha_{\rm MLT}$ (we notice that the YY01 models are computed accounting for He-diffusion, and also their 
solar calibration of $\alpha_{\rm MLT}$ is obtained considering this process).
The evolving mass is the same to within 0.02$M_{\odot}$, a  
difference which does not influence at all the RGB location. The P00  
models are computed with an amount of overshooting from the formal  
boundary of the convective envelope, but this does not affect the RGB   
$T_{\rm eff}$.  
  
The V00 and SW98 models are basically identical, the  
P00 ones are systematically hotter by $\sim$200 K, while the YY01 ones   
have a different shape.  
This comparison shows clearly that if one set of MLT solar calibrated RGBs  
can reproduce a set of empirical RGB temperatures, the others cannot,  
and therefore in some case a solar calibrated $\alpha_{\rm MLT}$  
value may not be adequate.  
The reason for these discrepancies must be due to some    
difference in the input physics, like the EOS and/or the boundary conditions,    
which is not compensated by the solar recalibration of $\alpha_{\rm MLT}$.   
   
\begin{figure}   
\plotone{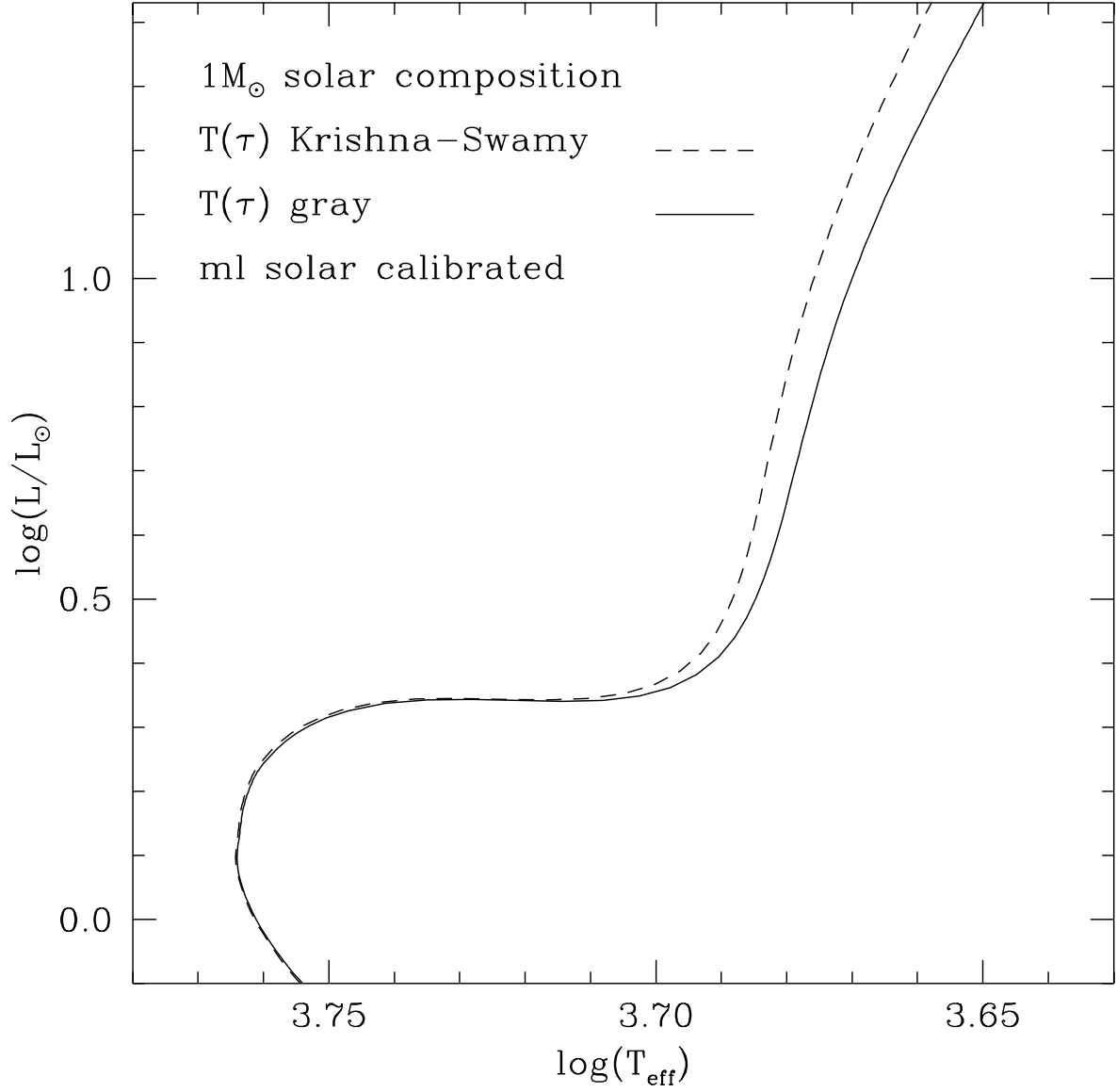}   
\caption{Two models for a 1$M_{\odot}$ star, solar chemical   
composition and $\alpha_{\rm MLT}$, and two different T($\tau$) relationships.   
\label{ttau}}   
\end{figure}

To illustrate this point   
in more detail, Fig.~\ref{ttau} shows two evolutionary    
models for the Sun,   
pushed up to the RGB. The only difference between them   
is the treatment of the boundary   
conditions. Two different T($\tau$) relationships,    
namely a gray one and the Krishna-Swamy~(1966) one have been employed.   
The value of $\alpha_{\rm MLT}$ for the two models in Fig.~\ref{ttau} has been    
calibrated in each case, in order to reproduce the Sun,    
and in fact the two tracks    
completely overlap along the main sequence, but the RGBs show a
difference of the order of 100 K 
\footnote{In this comparison the relative position of the RGBs  
computed with the two different T($\tau$) relationships is inverted  
with respect to the results in Fig.~\ref{modatm}. This comes from the  
fact that in Fig.~\ref{modatm} the models were computed at constant  
$\alpha_{\rm MLT}$, while in Fig.~\ref{ttau} $\alpha_{\rm MLT}$ has been recalibrated in order to  
match the Sun. The solar $\alpha_{\rm MLT}$ obtained in case of the gray T($\tau$)
is 1.62 (using the MLT formalism in Cox \& Giuli~1968), 
while the Krishna-Swamy one provides 1.82; due to the  
higher sensitivity of the RGB to $\alpha_{\rm MLT}$, the location of the two RGBs 
is reversed.}.     
  
This fact alone does not explain the differences  
among the RGBs in Fig.~\ref{rgbcomp} --   
the P00 (and YY01) models employ a gray T($\tau$)   
while the SW98 and V00 ones a Krishna-Swamy~(1966) T($\tau$), and the  
differences go in the opposite direction with respect to the effect  
shown in Fig.~\ref{ttau} -- but clearly points out the fact that one  
cannot expect the same RGB $T_{\rm eff}$ from solar calibrated models  
not employing exactly the same input physics. The logical conclusion  
is that it is always necessary to compare RGB models with observations  
to ensure the proper calibration of $\alpha_{\rm MLT}$ for RGB stars.  
  
Before concluding this section, we recall that   
there exist also extensive RGB computations    
(Mazzitelli, D'Antona \& Caloi~1995, Silvestri et al.~1998) using   
an alternative formalism for the computation of the superadiabatic   
gradient, which in principle does not require the calibration of any free   
parameter. It is the so-called Full-Spectrum-Turbulence theory (FST --   
see, e.g., Canuto \& Mazzitelli~1991, Canuto, Goldman \&   
Mazzitelli~1996), a MLT-like formalism with a more sophisticated   
expression for the convective flux, and the scale-length   
of the convective motion fixed a priori (at each point in a convective region,   
it is equal to the harmonic average   
between the distances from the top and the bottom convective   
boundaries). As far as the RGB   
$T_{\rm eff}$ is concerned, the FST predicts significantly different   
temperatures from a solar-calibrated MLT, at least for metallicities    
above Z$\sim$0.001.   
   
In Fig.~\ref{rgbcomp} an RGB computed with the FST (Silvestri et   
al.~1998), Z=0.001 and t=12 Gyr is also displayed. The FST   
computation adopted a gray T($\tau$) relationship as surface boundary   
condition, and the same opacities as the displayed MLT models.   
It is interesting to notice that, at least at this metallicity, the   
FST model is located within the temperature   
range spanned by the various solar calibrated $\alpha_{\rm MLT}$ tracks.   
   
Since there are basically no adjustable parameters in the FST formalism,   
any mismatch between predicted and observed temperatures for RGBs in GCs   
could still be due to systematic errors in the adopted $T_{\rm eff}$
and metallicity scale, and/or to inaccuracies of other aspects of the input physics, which,   
in case of the MLT, can be masked by a suitable calibration of $\alpha_{\rm MLT}$.    
It is therefore fair to say that   
a stringent test about the adequacy of the FST for determining the RGB    
$T_{\rm eff}$ has to wait for a final assessment of not only    
the RGB empirical temperature (and metallicity) scale problem,    
but also of all physical inputs affecting the    
model $T_{\rm eff}$ (e.g., EOS, boundary conditions).    
   
\subsection{Color transformations and bolometric corrections}   
   
As previously mentioned, one needs bolometric corrections and color   
transformations in order to perform comparisons between theory and observations.    
After a set of theoretical models is computed, there   
is a vast selection of available transformations that can be applied to   
the theoretical results; they can be divided into three categories.   
   
\noindent   
i) Results from theoretical model atmospheres.   
   
\noindent    
The data usually consist of tables providing bolometric corrections to
the $V$ band ($BC_{V}$),    
plus colors in  various photometric bands,  
as a function of chemical composition, surface gravity and $T_{\rm eff}$ . 
These are purely theoretical results, based on the computation of large sets of   
model atmospheres and spectra.   
The latest widely available sets covering RGB models for a large range of   
chemical compositions are from Kurucz~(1992),    
Buser \& Kurucz~(1992), Castelli, Gratton \& Kurucz~(1997),    
Castelli~(1999).   
   
\noindent   
ii) Results from theoretical model atmospheres, recalibrated on empirical   
data.   
   
\noindent   
As before, but this time the theoretical colors and $BC_{V}$ values have    
been modified (the procedure varies according to the author) in order to    
reproduce selected observational constraints. Examples of these transformations   
are the tabulations by Green~(1988), Lejeune, Cuisinier \& Buser~(1998),   
Houdashelt, Bell \& Sweigart~(2000).   
   
\noindent   
iii) Empirical transformations.   
   
\noindent   
They consist of relationships between   
empirically determined $T_{\rm eff}$, colors and $BC_{V}$   
at different metallicities, which are provided in term of the evolutionary phase.    
Alonso et al.~(1996, 1999) results are the most extensive   
ones, covering both Main Sequence and RGB stars of spectral type F0-K5
in different photometric bands. Sekiguchi \& Fukugita~(2000) also have
derived an empirical $(B-V)$-$T_{\rm eff}$ relationship for the same
class of stars. 
Empirical results for RGB stars are also provided by Montegriffo et   
al.~(1998) and von Braun et al.~(1998).   
A severe limitation of this approach is that the range of chemical   
compositions and evolutionary phases covered by these transformations   
is strongly constrained by the evolutionary stage and chemical composition    
of the calibrating stars.  
   
Due to the arbitrary zero-point of the $BC_{V}$ scale (e.g., Castelli~1999),   
when computing $V$ magnitudes from the values of the bolometric luminosity   
provided by the evolutionary tracks, and  
$BC_{V}$ given by the selected sets of transformations, one has first to adjust   
the bolometric magnitude of the Sun in order to reproduce    
the solar visual luminosity ($\rm M_{\sl V, \odot}$=4.82$\pm$0.02, Hayes~1985)   
with the provided solar $BC_{V}$ (see, e.g., the discussion by Castelli~1999).   
An equivalent possibility is to fix the value of the bolometric magnitude of the   
Sun and adjust the zero point of the $BC_{V}$ scale in order to reproduce    
$M_{V, \odot}$. With this calibration of the solar bolometric  
magnitude (or of the $BC_{V}$ zero point) one can then consistently   
derive the magnitudes in other photometric bands -- e.g. $\rm M_{I}$,  
$\rm M_{K}$ --   
from the computed $V$ magnitude and the appropriate color  
index, e.g. $\rm M_{I}$=$\rm M_{V}$$-(V-I)$, $\rm M_{K}$=$\rm M_{V}$$-(V-K)$.  
It is also worth remembering that the zero point of the  
color indices predicted by model atmospheres computations   
is usually set by reproducing the colors of Vega (e.g. Castelli~1999).  
  
After properly calibrating the zero point of the $BC_{V}$ scale,  
one finds that the various available sets of transformations,  
when applied to a given set of stellar   
models, provide $\rm M_{V}$ values within $\sim$0.05 mag    
(see also Weiss \& Salaris~1999).     
Along the RGB the derivative $\frac{\Delta color}{\Delta   
\rm M_{V}}$ is very small, therefore this uncertainty is not particularly   
relevant when comparing the position of theoretical RGB models with   
observations. However, the same kind of uncertainty on color transformations   
has much more dramatic consequences.    
   
\begin{figure}   
\plotone{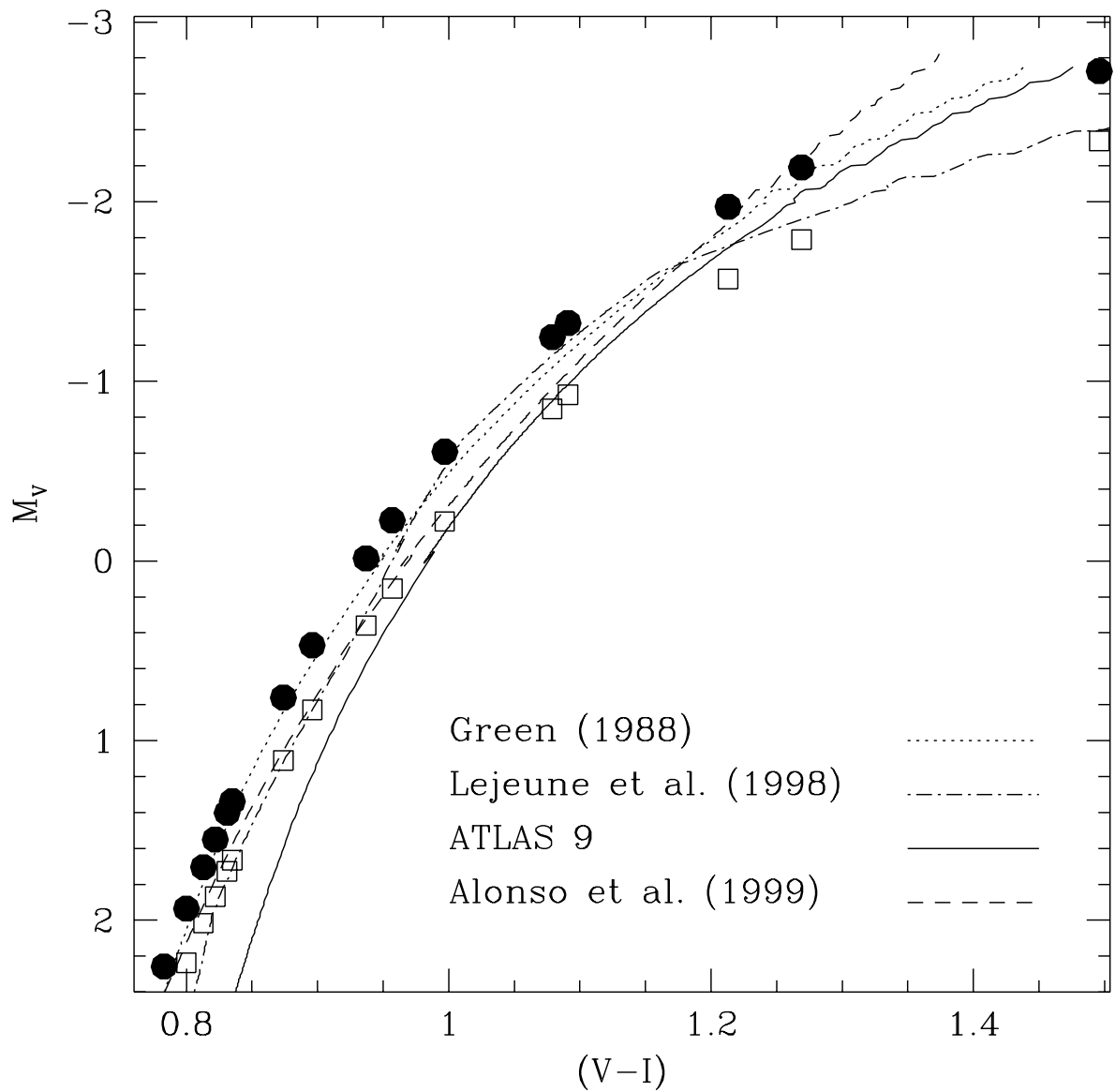}   
\caption{An isochrone with t=10 Gyr and Z=0.001 (from SW98)    
transformed to the $V-(V-I)$ plane using the 4 sets of   
transformations discussed in the text. Filled circles and empty squares
denote the RGB location (from Saviane et al.~2000) of Galactic GCs 
with this metallicity (assuming the same [$\alpha$/Fe] as in the SW98
models), considering both the Carretta \& Gratton~(1997 -- filled circles) 
and Zinn \& West~(1984 -- empty squares) [Fe/H] scales (see text for
more details).  
\label{colcomp}}   
\end{figure}

Let us consider, as an example, the $V-(V-I)$ Johnson-Cousins   
CMD (the situation in other photometric bands is similar, if not worse).    
Figure~\ref{colcomp} displays an isochrone with Z=0.001 and t=10   
Gyr (from Salaris \& Weiss~1998) using 4 different sets of   
transformations, namely, the ATLAS~9 theoretical transformations from   
Castelli, Gratton \& Kurucz~(1997), the semiempirical ones by    
Green~(1988) and Lejeune et al.~(1998),    
and the empirical ones by Alonso et al.~(1999),   
based on the IRFM temperature scale.   
For the sake of comparison, filled circles and empty squares 
denote the RGB location
of Galactic GCs with this metallicity (assuming the same [$\alpha$/Fe]
enhancement as in the Salaris \& Weiss~1998 models), taken from
the analytical relationships by Saviane et al.~(2000) for the
HB distance scale by Lee, Demarque \& Zinn~(1990 -- choices
of the distance scale differing by $\sim \pm$0.10 mag around this
value do not appreciably modify the comparison) and, respectively, 
the Carretta \& Gratton~(1997) [Fe/H] scale (filled circles) and the
Zinn \& West~(1984) one (empty squares). 

The differences in the RGB location and shape of the theoretical
models are basically due to the differences in the   
color-$T_{\rm eff}$ relationships. At a fixed value of $\rm M_{\rm   
V}$, $(V-I)$ color differences span a range between $\sim$0.05 and   
$\sim$0.10 magnitudes.   
This range of colors implies an uncertainty   
of $\sim 0.3-0.5$ dex when determining [Fe/H] from the comparison   
of observed and theoretical RGB CMDs.   
If the color difference between Turn-Off (TO) and base of the RGB   
is employed to estimate GC absolute ages, the uncertainty on the   
derived age would be again very high (assuming the main sequence  
color is unchanged), since a variation by $\sim   
0.01-0.02$ mag of this color difference   
corresponds to an age variation by about 1 Gyr.    
  
\section{The CMD as a tool for calibrating RGB models}   
   
A comparison between the observed and predicted CMD location of RGB   
models allows one in principle to test the reliability of the physical    
inputs employed   
to compute the structure of the stellar external layers. This is because, as   
previously discussed, the CMD location of RGB stars depends on the   
description of the envelope superadiabatic region (low-T opacities and   
convection treatment), the envelope EOS, the surface boundary   
conditions and color transformations adopted.   
   
Da Costa \& Armandroff~(1990) and more recently  
Saviane et al.~(2000) and Ferraro et al.~(2000) have published very   
useful relationships   
between colors and morphology of the RGB as a function   
of [Fe/H], from multicolor observations of a sample of galactic GCs.   
These relationships provide a test-bench for theoretical models, even   
though they depend on the adopted [Fe/H] scale (still   
somewhat uncertain, as already mentioned), cluster reddenings    
and, to a lesser extent, on the assumed distance scale.   
The general trends with respect to the stellar metallicity (change of   
location and shape of the RGB, and their dependence on the wavelength   
bands employed) predicted by the models are in broad qualitative agreement    
with observations, but the empirical relationships are more easily used   
to calibrate the RGB models rather than testing their reliability.   
Because of the still relevant uncertainties in the predicted CMD location   
of RGB models -- compare, e.g., the range of shapes, $T_{\rm eff}$   
and colors attainable by selecting various combinations of the    
different theoretical models and color transformations   
displayed in Figs.~\ref{rgbcomp} and ~\ref{colcomp} -- a comparison of   
theory with observations can be used to calibrate the mixing length   
and constrain the color transformations adopted in the model computations.   
This is precisely the approach followed by, e.g., Weiss \& Salaris~(1999)   
and V00.  
  
\section{Surface abundances and mixing mechanisms along the RGB}   
  
The standard notion of element abundance variations in GC stars is that,  
within the errors, there are no intra-cluster  
variations. To first order, and with only a small number of exceptions  
($\omega$~Cen, and possibly M22), this seems to be correct.  
Looking at detailed individual abundances, however, this is no longer  
true. As a consequence of the first dredge-up, standard RGB models predict a  
modest change in elements participating in proton-capture  
nucleosynthesis (\S~2.1). Since  
the depth of convective mixing extends to  layers with $T \approx  
2\cdot10^6$~K at most, only helium, lithium, carbon and nitrogen and their isotopes  
are concerned. In particular, the $\Iso{12}{C}/\Iso{13}{C}$ ratio is  
predicted to drop from the initial cosmic value of 90 to about 25, but  
not to that of CN-equilibrium, which is of order 3. These variations  
are predicted to take place near the base of the RGB in
low- and intermediate mass stars, and agree with observations.  
  
This canonical picture about RGB abundances, however, is challenged by an  
increasing amount of observational data demonstrating its limited  
validity. In the following, we will summarize this observational  
material and discuss possible theoretical explanations.  
In the context of this review, these observations clearly demonstrate  
that the physical processes taken into account in canonical stellar  
evolution calculations are insufficient and that the theory needs to  
be extended.  
The reader may find additional material in the excellent reviews by  
Kraft~(1994) and Da Costa~(1998), although the amount of  
observations has grown a lot since their publication.  
  
\subsection{The observational evidence}  
  
\subsubsection{Metal-poor field stars}  
\label{dm:field}  
  
There is now large evidence for mixing beyond the canonical first  
dredge-up prediction in metal-poor field stars. Charbonnel \& do  
Nascimento~(1998) summarized the literature data to find that 
in 95\% of the low-mass metal-poor field stars the  
$\Iso{12}{C}/\Iso{13}{C}$ ratio is lower than expected. The average  
value is of order 10, but a few stars also reach   
$\Iso{12}{C}/\Iso{13}{C} \leq 6$. Charbonnel \& do  
Nascimento~(1998) identify two  
interesting trends: firstly, there is a clear anti-correlation between  
isotope ratio and brightness, setting in at $M_V < 2$, i.e.\ approximately  
at the level of the RGB bump, and secondly, the lowest  
$\Iso{12}{C}/\Iso{13}{C}$ values are reached by those stars showing  
the highest surface rotation, $v\sin i$. The first fact provides  
substantial evidence that the isotope anomaly is due to intrinsic  
extra-mixing beyond the established first dredge-up during  RGB  
evolution; the second one hints that there is a correlation   
with rotation, which might possibly initiate this additional  
mixing.   
  
Gratton et al.~(2000) confirmed these results by a dedicated abundance  
analysis of 62 low-mass metal-poor field stars with HIPPARCOS parallaxes. They confirmed the  
trends found earlier and added additional details. They conclude that  
a second mixing episode sets in after the bump, resulting in lowered  
carbon abundances and carbon isotope ratios (but ``distinctly higher  
than the equilibrium value''), an increased nitrogen  and a  
declining Li abundance down to almost zero. They confirm that this  
happens in basically all stars of their own sample and 43 additional  
stars from the literature. The anticorrelation with brightness is  
evident, too. Another very important result is that neither O nor Na  
show any abundance variation in these field stars. This is, as we  
shall see below, quite different from similar stars in globular  
clusters.   
  
Most recently, Keller, Pilachowski \& Sneden~(2001) confirmed the   
carbon isotope results  
from IR-data. From a sample of literature  
data they also find that the amount of mixing seems to decline with  
increasing metallicity.   
  
\subsubsection{Globular cluster stars}  
\label{dm:cluster}  
  
The quantity and  quality of observations and the rate of new data  
acquisition is   
impressive. In the following, only a few highlights illustrating the  
main trends can be mentioned. For more extended references to  
observational material the reader should consult the current literature.  
  
Carbon abundance variations along GC RGBs have been detected as early  
as 1979 in M92 (Bell, Dickens \& Gustafsson~1979), the cluster which remains to be one of  
the most important cases displaying evolutionary effects  
(Carbon et al.~1982, Langer et al.~1986, Bellman et al.~2001).   
The variations are in agreement with those  
in field stars, i.e.\ a declining carbon abundance and isotope ratio, and  
an increasing nitrogen abundance with brightness.   
Similar cases include M15 (Trefzger et al.~1983), NGC~6397  
(Briley et al.~1990). In addition, star-to-star variations in O and Na were  
detected in a significant number of clusters, most prominently in  
NGC~6752 (Suntzeff \& Smith~1991), NGC~3201 (Gonzalez \&  
Wallerstein~1998),   
M3 (Suntzeff~1981) and M13  
(Suntzeff~1981, Pilachowski et al.~1996), the latter cluster exhibiting the most extreme  
spread in abundances. $\omega$~Cen (Norris \& Da Costa~1995), the  
only cluster with a clear intra-cluster iron abundance (or total  
metallicity) spread, also displays the full spectrum of   
anomalies. Oxygen tends to be underabundant by up to 0.5 to 1.0~dex with respect  
to the standard, $\alpha$-enhanced metal composition ([O/Fe]$\sim  
0.4$), while sodium is overabundant up to more than 0.5~dex.  
The important point is that the two elements are anticorrelated  
in their abundance variations. There is -- in contrast to the  
CNO-elements -- no clear evidence for a development of the anomalies  
along the RGB, although in M13 they seem to be absent below the bump,  
but fully developed above it. Most importantly, and as mentioned  
above, the O-Na-anticorrelation or any other anomaly in their  
abundances is definitely absent in field stars (Gratton et  
al.~2000). This points to the influence of some environmental effect.   
  
Many of the best observations have been obtained with the KECK/HIRES  
instrument, although smaller telescopes are sufficient for  
spectroscopy of the brightest stars. Since the year 2000, the VLT/UVES  
instrument has become available, allowing the observations of cluster  
stars at the subgiant branch (SGB)  
or TO. First results have been published already by Castilho et al.~(2000) and    
Gratton et al.~(2001) about NGC~6397 and NGC~6752. While for the first  
cluster the comparison between TO- and SGB-stars shows no significant  
abundance differences and a very low scatter in oxygen, NGC~6752  
already displays the full range of the O-Na-anticorrelation close to  
the main sequence. These observations provide strong evidence for a  
non-evolutionary source for the abundance anomalies, as do the results  
by King et al.~(1998) about M92, in which the corresponding Na and Mg (see  
below) anomalies are claimed to be found in subgiants, too.  
  
More metal-rich clusters ([Fe/H]$ \geq -1.4$) support this:  
in a number of clusters (47~Tuc, Cannon et al.~1998; M4, Ivans et al.~1999;  
M71, Briley \& Cohen~2001)  a bimodal  
C-N abundance distribution is present all along the  
cluster sequence, or -- as in M5 (Ivans et al.~2001) -- the extent of the  
anomaly may even be anti-correlated with evolutionary state. All these facts  
are interpreted in terms of the presence of a primordial source such  
as contamination by more massive stars.  
  
The final point in favor of a primordial component responsible for at  
least some   
abundance variations is the fact that also a Mg-Al-anticorrelation has  
been found in several clusters (among them $\omega$~Cen,   
Norris \& Da Costa~1995; M13,  
Shetrone~1996a; M4, Ivans et al.~1999), where Al can be overabundant by  
up to 1~dex. As will be discussed below, the conversion of $^{24}$Mg  
to $^{27}$Al is possible only in stars more evolved than the RGB,  
since temperatures of $\approx 7 \cdot 10^7$K are needed  
(Langer, Hoffman \& Zaidins~1997).   
Al-production from other Mg-isotopes fails to explain the high  
Al-enrichment and leads to conflicts with the Na-abundance trends.  
The high Al abundances found in cluster stars are probably the most  
difficult anomaly to be explained.  
  
\subsection{Proton-capture nucleosynthesis}  
\label{dm:proton}  
  
The result of the standard first dredge-up, taking place on the   
SGB and lower RGB is to transport material that has participated in  
hydrogen-burning into the convective envelope, mixing it there with  
matter of the initial composition. The observed abundances of  
CNO-elements are in complete agreement with the theoretical  
predictions, and in particular show the constancy of the sum of C, N,  
and O, the proper correlations, and isotope ratios. They can be  
explained, thus, by mixing of proton-capture nucleosynthesis products.  
  
In addition to the well-known CNO-cycle, analogous reaction cycles involving  
heavier elements exist at higher temperature. In Fig.~\ref{f1} we  
show the NeNa- and MgAl-cycles and in Fig.~\ref{f2} the abundance  
profiles in and above the hydrogen-burning shell of a typical  
RGB cluster star. It is obvious that abundance variations due to first  
proton-capture reactions, such as the C$\rightarrow$N conversion  
appear outside the region of a significantly reduced H-abundance and  
thus of a varying molecular weight.  
  
The elements affected (C, N, O, Ne, Na, Mg, Al) are exactly those
which show 
star-to-star  
variations in clusters, providing the first evidence that the  
anomalies are the result of proton-capture reactions. 
Denissenkov \& Denissenkova~(1990)  
demonstrated that the O-Na-anticorrelation can be explained self-consistently  
by the assumption that both elements have been subject to nuclear  
processing in a typical RGB hydrogen-shell. The crucial point is  
that Na and O are not participating in the same cycle (as is the  
case for the C-N-anticorrelation), but in the NeNa- and CNO-cycle,  
respectively. However, within the shell oxygen is depleted  and  
sodium enhanced at the same location, and hence   
temperature (see Fig.~\ref{f2}; the first 
rise at $0.08 \lesssim \delta m \lesssim 0.13$ of  
the model shown). Note that in the outermost layers of this region Na is enhanced  
without significant depletion of O; such cases are observed,  
too. Unfortunately, the accompanying   
Ne-depletion cannot be detected. Langer, Hoffman \& Sneden~(1993) and others confirmed  
that a typical RGB H-shell could be the source for the anomalies  
up to Na, and since then the observations are being explained in terms  
of proton-capture nucleosynthesis. The most up-to-date reaction rates  
as well as a discussion of their errors can be found in  
Arnould, Goriely \& Jorissen~(1999).   
  
It is of special interest that the stable aluminium isotope  
$\Iso{27}{Al}$ is produced only in the very deep interior of the  
H-shell and at the expense of $\Iso{25}{Mg}$ (Fig.~\ref{f2}, where the  
initial magnesium abundance was raised to $\mathrm{[^{25}Mg/Fe]}=1.2$; see below).   
Langer \& Hoffman~(1995) investigated in detail the nucleosynthesis of Na, Al and  
Mg. Their results can be summarized as follows: (i) under RGB-typical  
conditions Al is produced almost exclusively at the expense of  
$\Iso{25}{Mg}$; for isotope and element ratios typical of field halo  
stars this results  
in only a modest Al enrichment of $\approx 0.3$~dex; (ii) to reach the  
high [Al/Fe]-levels observed, $\mathrm{[^{25}Mg/Fe]} \gtrsim 1.0$ has  
to be assumed; (iii) the conversion of $\Iso{25}{Mg}$ into  
$\Iso{27}{Al}$ leads to a depletion of only 0.2~dex in Mg; (iv)  
significant Al-production happens only at very high temperatures; as a  
consequence, helium-enriched material is mixed into the  
envelope. The second of these points requires an additional primordial component  
and the first one is in contradiction with the fact that it is  
$\Iso{24}{Mg}$, which is depleted in M13  
(Shetrone~1996b, but see Ivans et al.~1999 for a different result on  
M4). And finally, the depletion of Mg in M13 is as large   
as 0.4~dex, which can be reconciled with (iii) only with  
difficulties.   
A further complication arises from the fact that nucleosynthesis  
calculations show that Na has a higher overproduction than Al  
(Langer et al.~1993), while observations (Norris \& Da Costa~1995) indicate the  
opposite. Langer et al.~(1997) showed that   
a better agreement with all element abundances can be reached if the  
burning temperature of the hydrogen shell is assumed to be around  
$7 \cdot 10^7$~K, more typical for asymptotic giant branch (AGB)  
stars. In RGB stars, $5.5\cdot  
10^7$~K is the maximum value reached by very metal-poor stars close  
to TRGB, while Al is overabundant at much lower brightness as well.  
  
\subsection{Theoretical models for the abundance anomalies}  
  
\subsubsection{Evolutionary scenario}  
\label{dm:evo}  
  
The very clear evidence from field stars (\S~\ref{dm:field}) and  
CN-variations in some clusters (\S~\ref{dm:cluster}) requires  
that at least part of the complete explanation for all the anomalies  
discussed is to be found in a non-canonical effect taking  
place during   
the RGB-phase, due to which the products of nucleosynthesis in the red  
giant's own H-burning shell appear at the surface. This  
requires some additional mixing between shell and the bottom of  
the convective envelope. The mixing must be rather slow, otherwise the  
trends with luminosity would not be observed and the whole envelope  
could probably be processed by the shell. Most of the observations  
supporting this evolutionary scenario indicate that the extra mixing  
sets in at or after the bump, i.e.\ the point at which there is an  
almost vanishing difference in molecular weight between the outer  
shell regions ($\delta m \gtrsim 0.10$ in Fig.~\ref{f2}) and the  
convective envelope. However, in M92 the carbon depletion might start  
even earlier than that (see Bellman et al.~2001).   
  
Evolutionary models, pioneered by Sweigart \& Mengel~(1979), therefore try to  
reproduce the observations by adding proton-capture nucleosynthesis in  
the NeNa- and MgAl-cycles and some kind of extra-mixing to the  
canonical stellar evolution calculations. Different implementations  
ranging from \lq{conveyor belt}\rq\ approaches (Wasserburg, Boothroyd \&  
Sackmann~1995) to models  
fully incorporating diffusive mixing (Charbonnel~1995, Weiss,  
Denissenkov \& Charbonnel~2000b) exist,  
but none is -- up to the present time -- completely self-consistent, in the  
sense that the physical reason for the mixing (probably differential  
rotation), its extent and speed, the full nucleosynthesis, and the  
back-reaction on the model are solved simultaneously. In the following  
we will describe a two-stage approach used by Denissenkov \& Weiss  
(Denissenkov \& Weiss~1996, and later work) and -- in a slightly variant form -- by  
Cavallo, Sweigart \& Bell~(1998, and follow-up work).  
  
This method separates the canonical from the non-canonical aspects of  
the problem. A stellar model of appropriate mass (typically  
$0.8\,M_\odot$) and composition is evolved within the canonical   
framework along the RGB. Several models at various epochs are then used  
for the background structure ($M_r(r)$, $T(r)$, $P(r)$, $L_r(r)$) on  
which the extra-mixing and detailed nucleosynthesis are performed.   
The background structure of the region between shell and convective  
envelope can be obtained for any given timestep by  
interpolation in the relative coordinate $\delta m = (m_r -  
m_{bs}) / (m_{bcz} - m_{bs})$, where  
$m_{bs}$ and $m_{bcz}$ are the relative mass coordinates  
of the bottom of shell and convective zone, respectively  
(Denissenkov \& Weiss~1996). Nucleosynthesis then is calculated with this background  
physical structure, by using an  
extensive network of up to 26 nuclei and 69 reactions  
(Denissenkov et al.~1998). The mixing is done by solving a diffusion equation  
with diffusive speed (the diffusion constant) and  
penetration depth as free parameters. Both are identical for all elements.  
The advantage of such an approach is that it  
allows the calculation of a range of extra-mixing parameters and an  
economical exploration of stellar metallicity (Cavallo et al.~1998).  
The disadvantage is that it has to assume that the mixing has no  
influence on the   
background structure evolution, i.e.\ that there is no significant  
back-reaction. Obviously,  
the approach is always inconsistent as the mixing of the various  
metals has not been taken into account in the stellar model  
calculations.   
As long as no change in the H/He-profile takes   
place and thus the mixing does not touch regions of declining  
hydrogen abundance in the shell ($\delta m \lesssim 0.08$), the hope  
is that the inconsistency does not pose a severe problem. However, it  
also implies that no very deep mixing can be investigated.  
  
All results confirm that CN- and  
ONa-abundance anomalies can be reproduced by the deep-mixing scenario  
with reasonable mixing assumptions  
(e.g., Langer et al.~1993, Charbonnel~1995, Denissenkov \& Weiss~1996,  
Cavallo et al.~1998, Denissenkov \& Tout~2001). In a number of papers, these  
assumptions are based on a physical model for rotation-induced mixing.  
For example, the results by Denissenkov \& Tout~(2001), repeated in Fig.~\ref{f3},  
were obtained this way. Due to the physical approach the   
derived effective diffusion constant varies with depth, and is of  
order $10^8 - 10^9\,\mathrm{cm}^2\,\mathrm{s}^{-1}$. The diffusive  
speed drops toward zero at $\delta m \approx 0.08$; this point corresponds  
to the mixing depth used in Denissenkov \& Weiss~(1996) and Weiss et  
al.~(2000b).   
These numbers are in good agreement with the parameters used in the two  
papers quoted. The penetration depths are actually shallow enough  
to justify the two-stage approach.   
Figs.~\ref{f3} and \ref{f4} illustrate the successful reproduction of  
observed anomalies in several clusters.  
  
As a (rather important) side-effect, the additional mixing also  
ensures that most of $\Iso{3}{He}$ produced in the   
pp-I-chain on the main-sequence is destroyed, thereby resolving  
one of the problems of galactic chemical evolution -- that of  
$\Iso{3}{He}$-overproduction (see Charbonnel \& do Nascimento~1998,  
Weiss et al.~1996, and references therein).  
  
The calculations based on rotation-induced mixing  
(Charbonnel~1995, Denissenkov \& Tout~2001) also predict that the
mixing 
is not able to  
penetrate significant molecular weight changes (Charbonnel, 
Brown \& Wallerstein~1998),  
supporting the picture that the evolutionary effects can start only at  
or after the bump and that no significant amounts of helium will be  
mixed to the surface. The same models predict that no additional  
mixing takes place in stars of $M\gtrsim 2.0\,M_\odot$ and that the  
degree of extra-mixing diminishes with increasing  
metallicity (Cavallo et al.~1998). This is in agreement with observations  
(e.g. Gilroy~1989 and Briley, Smith \& Claver~2001).   
  
In contrast to this, Sweigart(1997a,b), argued for a  
penetration of the molecular weight barrier and resulting helium  
mixing accompanying   
extreme O-Na-anomalies such as those found in M13.   
Such \lq{very deep mixing}\rq\ would have significant bearing for both the  
evolution along the RGB, the brightness of the TRGB, the HB  
brightness level and the HB morphology. Some of these predictions will  
be discussed further in \S~8 and \S~9.1.  
Weiss et al.~(2000b) investigated the consequences on the  
proton-capture element abundances with models in which the helium  
mixing itself was   
included into the full stellar evolution calculations (to account for  
the structural changes) and concluded that noticeable helium mixing  
would be accompanied by extreme O and Na changes (resulting from the  
second Na rise in Fig.~\ref{f2}) exceeding the observed  
levels. Therefore, based purely on nucleosynthesis arguments, the  
mixing of regions with $\triangle X \gtrsim 0.1$ could be excluded; on  
the other hand, Weiss et al.~(2000b) also found models which evolved to  
very high luminosities without strong helium mixing into the envelope  
and without violating the abundance observations. However, these were  
preliminary results to be investigated further with more physical  
mixing models.   
  
The \lq{normal deep mixing}\rq\ models all fail to explain the Mg-Al  
anticorrelation, because of   
the temperatures in the shell regions affected by the deep mixing  
being too low for Al production from $\Iso{24}{Mg}$.  
Fujimoto, Aikawa \& Kato~(1999), and Aikawa,  
Fujimoto \& Kato~(2001) proposed a variant  
of the deep mixing scenario, where the mixing repeatedly penetrates the  
whole H-shell, mixing protons into the hot helium-core, leading to  
shell flashes, in which also Al is synthesized.   
This scenario is in fact very similar to the helium-flash in Pop~III  
stars discussed in \S~2.3.  
While this model  
succeeds in reproducing several abundance trends, which remain  
unexplained in the simple deep-mixing scenario, the models are simple   
box-models (i.e.\ no complete stellar models) so far, and a physical  
explanation for these extreme mixing events is even more difficult.  
  
Concerning the Al problem, Langer \& Hoffman~(1995) noted that  
significant amounts of   
the radioactively unstable isotope $\Iso{26}{Al}$ (lifetime: $\approx  
10^6$~years) can be produced in the same shell regions which are  
responsible for the   
O-Na-anticorrelation. Denissenkov \& Weiss~(2001) investigated this in detail with  
the two-stage method and demonstrated that under the assumption that the  
observed Al is actually $\Iso{26}{Al}$, all known correlations in  
$\omega$~Cen and M4 (Fig.~\ref{f4}) can be simultaneously explained with standard  
mixing parameters. In fact, the mixing speed is high enough as  
compared to the isotope's lifetime  that   
enough $\Iso{26}{Al}$ is always transported to the surface, replacing the  
decayed atoms. If the Al could be confirmed as $\Iso{26}{Al}$, either by  
isotopic line shifts or the emitted 1.8~MeV $\gamma$-line (both not  
being possible as yet), this would be a direct proof for the on-going  
mixing process. As mentioned earlier, these models also needed a   
primordial enrichment of $\Iso{25}{Mg}$ by a factor 10.  
  
\subsubsection{Primordial scenarios}  
  
As was shown in \S~\ref{dm:cluster} and \S~\ref{dm:proton} there  
are a number of arguments in favor of a primordial component: These  
are the failure to produce the large Al-overabundances in standard  
RGB H-shells consistently with the Mg and Na data, and the  
finding of anomalies in a number of cluster stars already on the SGB,  
or even earlier. Also, Ivans et al.~(1999) argue that in M4 there exists a  
``primordial floor'' of abundance anomalies (notably in Al), on which  
an evolutionary component is superimposed.  
  
Most papers employing primordial components turn to AGB-stars of  
higher mass, in which proton-nucleosynthesis takes place at higher  
temperatures. We refer the reader to Denissenkov, Weiss \&  
Wagenhuber~(1997), Denissenkov et al.~(1998) and Ventura et al.~(2001);  
the latter is focused on the so-called hot-bottom burning  
phase. This approach explains in particular the high Al and low Mg (probably  
$\Iso{24}{Mg}$) observed in two subgiants in NGC~6752  
(Gratton et al.~2001). Since in the primordial picture the nucleosynthesis  
and mass loss takes place in more advanced stages of evolution, we  
will not go into the details here.  
  
\subsection{Summary of RGB abundance anomalies and consequences}  
  
The existence of star-to-star variations in the abundances of  
proton-capture elements has been demonstrated without doubt. In fact,  
it appears as if they are a generic property of all low-mass ($M  
\lesssim 2.0\,M_\odot$) metal-poor stars, as long as they are  
restricted to CNO-elements. The same data strongly point to an  
intrinsic evolutionary process in red giants, which still has to be  
fully understood and incorporated into self-consistent models. Only  
cluster stars seem to exhibit the anomalies in heavier elements (Na,  
O, Mg, Al) as well. It is not yet clear, whether an evolutionary  
component is necessary to explain those, but a primordial one seems  
unavoidable. This will lead to interesting consequences for the  
intra-cluster gas chemical evolution.   
  
Although some reactions in the NeNa- and MgAl-cycles are still  
uncertain, this general picture will not depend on them. However,  
nuclear data would be of great help to narrow the range of conditions  
for proton-capture nucleosynthesis. And finally, abundance data as a  
function of luminosity along cluster sequences are necessary from the  
TO to the TRGB in order to investigate the contributions of  
primordial and evolutionary components. 
 
Before concluding this section, we notice that, 
as long as one tries to explain by deep mixing only 
the CNO and NeNa-cycle anomalies, and therefore no appreciable amount 
of hydrogen is brought into the shell (and helium dredged into the envelope), 
the evolutionary timescales and RGB location predicted by canonical models 
are not affected by this added element transport mechanism (see, e.g., 
Weiss et al.~2000b and discussion in \S 8).

\section{The LF as a tool for testing the inner structure of RGB models}   
   
The analysis of the RGB LF   
is of paramount importance in order to check the accuracy of    
the inner structure of theoretical RGB models. In particular,   
as discussed by Renzini \& Fusi-Pecci~(1988), through the LF one can     
test the chemical stratification inside the star.   
This follows from the fact that the hydrogen abundance   
sampled by the thin H-burning shell strongly affects the rate of   
evolution, hence the star counts along the RGB.   
When a statistically sizeable sample of RGB stars is available, one could expect    
to check the envelope chemical profile with a mass resolution of the same order    
of magnitude of the shell thickness.    
Moreover, the LF is unaffected by the existing uncertainties in the   
$T_{\rm eff}$ and color determination.   
   
Langer, Bolte \& Sandquist~(2000) have recently claimed the existence of a   
discrepancy between the theoretical and observed RGB LF of the two GCs, M5   
and M30 (see also Vandenberg et al.~1998) . They found an overabundance on stars in the upper RGB,   
arguing that the cause might be the action of some very deep mixing   
mechanism active during the   
RGB ascent. This deep mixing (see \S 7 for a more detailed discussion  
about this point) would be able to bring some    
hydrogen from the envelope into the H-burning shell (and therefore an extra    
amount of He in the envelope), causing a longer lifetime and an increase of the   
number of observed RGB stars.   
A recent analysis by Zoccali \& Piotto~(2000) has found however a good   
general agreement between various sets of theoretical LFs and   
observations of a large sample of GCs -- some of which show the 
abundance anomalies discussed in \S 7 --, at least below [Fe/H]$\sim   
-$0.7. Possible discrepancies at higher metallicity, in the upper part   
of the RGB, need to be confirmed with further observational data.   
The LF slope predicted by theoretical models and its    
independence of the initial stellar metallicity   
looks therefore confirmed, supporting    
the relative timescale along the RGB predicted by canonical models, 
and showing that deep mixing, if effective, does not appreciably 
affect the predictions of canonical RGB models.   
A similar conclusion has been reached by Rood et al.~(1999) in their   
study of the LF of M3.   
   
\subsection{The RGB bump brightness}   
   
As already mentioned, during the RGB   
evolution the H-burning shell crosses the chemical discontinuity left over by the convective   
envelope soon after the first dredge-up. When the shell encounters this discontinuity, matter in   
the shell expands and cools slightly, causing a sharp drop in the stellar surface luminosity.   
After a while, thermal equilibrium is achieved again and the stellar luminosity starts to   
monotonically increase up to the TRGB. As a consequence, the structure   
passes three times across a narrow magnitude interval. This occurrence   
produces the characteristic bump in the theoretical LF.   
   
Since its detection in 47Tuc, the RGB bump has become   
the crossroad of several theoretical and observational investigations (Fusi Pecci et al.    
1990, Alves \& Sarajedini 1999, Ferraro et al. 1999, Zoccali et al. 1999). In order to detect   
the bump a large sample of RGB stars is required -- $\sim 10^3$
stars in the upper four magnitudes of the 
RGB according to Fusi Pecci et al.~(1990)   
-- and until a few years ago it was detected only in metal-rich stars.   
In fact, increasing the cluster metallicity, the luminosity extension of the bump is larger   
and it is also shifted  to lower luminosity, in a more populated RGB portion; both effects   
work in the direction of making the bump detection easier. Only in this last decade,    
thanks to the HST capability of imaging the dense GC cores, and    
the availability of large field CCD detectors, has the RGB bump been detected   
also in metal-poor GCs.   
   
Since the position of the bump is related to the location    
inside the stellar structure of the H-discontinuity, the deeper the chemical discontinuity   
is located, the fainter the bump luminosity will be. A comparison of   
the theoretical bump level with observations allows a test of the   
predicted maximum extension of the convective envelope in RGB models.   
The parameter $\Delta V_{\rm HB}^{\rm Bump}= V_{Bump}-V_{HB}$   
-- the V-magnitude difference between the RGB bump and the HB   
at the RR Lyrae instability strip  level (Fusi Pecci et al. 1990,   
Cassisi \& Salaris 1997, hereinafter CS97)  -- is commonly adopted   
as a measure of the bump brightness. From the observational point of  
view it is a distance independent quantity, but   
theoretically \vhbb depends not only   
on the bump level predicted by models, but also on the HB luminosity   
set by the value of $M_{core}^{He}$ at the He-flash, which is still   
subject to the uncertainties discussed before.   
   
For a given set of input physics employed in the model computation   
\vhbb is affected mainly by the total metallicity [M/H]   
and the age of the cluster. As far    
as the age is concerned,   
an increase by 1 Gyr at a given metallicity   
causes an increase of \vhbb by about 0.024 mag. As the HB luminosity
is almost unaffected by   
the age  -- for typical GC ages -- this change is due to the    
change in the bump luminosity: when the cluster age increases, the   
mass of the evolving RGB star decreases, and the H-discontinuity
location deepens because of the larger   
convective regions.   
The effect associated with the cluster metallicity is significantly more relevant.   
For an age of 15 Gyr, a change by $+0.20$ dex of [M/H] increases \vhbb    
by $\sim 0.1-0.2$ mag (the precise value depending on the actual   
metallicity). This is a consequence of   
the fact that the larger the star metallicity, the larger is the mass   
extension of the convective envelopes (larger radiative opacities   
within the structure) and the dimmer is the bump level; at the same time,   
when the metallicity increases, the HB becomes fainter because of the smaller   
$M_{core}^{He}$ at the He-flash, but this brightness decrease is   
smaller than the change of the bump level, so that \vhbb increases.   
   
As a reference, we provide here the relationship derived from the   
canonical models by  CS97,   
for a cluster age equal to 15 Gyr and in the whole Galactic GCs metallicity range:   
   
\begin{equation}   
\Delta V_{\rm HB}^{\rm Bump}= 1.083 + 1.380[M/H] + 0.231[M/H]^2    
\end{equation}   
   
   
   
The first exhaustive comparison between theory and observations of   
\vhbb in galactic GCs has been performed by Fusi Pecci et   
al.~(1990). Using the theoretical models available at that   
time, they found that   
the observed dependence of \vhbb on cluster metallicity is in good agreement with theoretical   
predictions, but theoretical   
\vhbb values were smaller than the observed ones by $\approx 0.4$ mag.   
For a long time, this result has been considered a clear drawback of   
canonical theoretical   
models of low-mass RGB stars. It was therefore   
suggested by Alongi et al.~(1991) that this   
discrepancy can be removed accounting for the efficiency    
of overshooting from the base of the formal boundary of the   
convective envelope, which shifts the bump level to a lower   
brightness, because of the resulting deeper convective envelope.   
By parametrizing the mass extension of the overshooting region in units     
of the local pressure scale height ($H_{p}$), it was shown that    
an overshooting region of $\sim$0.7$H_{p}$ could resolve the discrepancy.   
   
More recently, CS97 have reanalyzed the problem using their updated   
canonical stellar models applied to a sample of 8 clusters with spectroscopical   
determinations of [Fe/H] and [$\alpha$/Fe], taking explicitly into   
account the effect of the $\alpha$-element overabundance.    
They concluded that their models (see previous relationship) provide a   
good match to observations.    
Zoccali et al. (1999) and Ferraro et al. (1999) have later produced   
observational \vhbb databases for, respectively, 28   
clusters obtained from HST observations -- all calibrated and reduced   
homogeneously -- and 40 clusters taken from literature data.   
In Fig.~\ref{bumpvhbb} (top panel), we show the comparison between   
models by CS97    
and the Zoccali et al.~(1999) data, plus additional clusters from    
the Ferraro et al.~(1999) catalogue.   
We have adopted the [Fe/H] scale provided by Carretta \& Gratton
(1997) 
and the labeled   
assumptions about the trend of the $\alpha-$element enhancement with
the 
iron abundances   
(see Zoccali et al. 1999 for a full discussion). We have considered an   
uncertainty on the global metallicity [M/H] of the order of 0.15 dex,  
which perhaps is a   
lower limit for the true uncertainties affecting both [Fe/H] and  
$[\alpha$/Fe] measurements   
(Carney 1996, Rutledge, Hesser \& Stetson 1997).   
   
The top panel of Fig.~\ref{bumpvhbb} shows a good agreement   
between theory and observations.   
In order to account for the current uncertainty on the zero point of the GC   
[Fe/H] scale, the bottom panel shows the same comparison  
employing the Zinn \& West~(1984) [Fe/H] scale. In this case, the   
agreement between theory and observations is worse, at least for   
metallicity lower than [M/H]$<-1.0$, where the two [Fe/H] scales are   
most different: the disagreement with theoretical predictions    
for a cluster age of 14 Gyrs is on average of about 0.20   
mag, a value which is almost a factor of 2 larger than the 
mean photometric uncertainty.   
   
Bergbusch \& Vandenberg~(2001) found a discrepancy of about 0.25 mag    
when comparing the results from their isochrones (V00 models)   
with \vhbb data for 4   
GCs at various metallicities (see their Figure~17).
The bump brightness from their models is in good agreement with the CS97   
results, while their HB models are about 0.05 mag fainter; the GC 
[Fe/H] values adopted in their comparison are within 0.1 dex of those
by Zinn \& West~(1984), whose metallicity scale they favored from
other considerations.
   
The conclusion to draw from this discussion is that the    
discrepancy found by Fusi Pecci et al.~(1990) may well be reduced to   
zero without the need to include overshooting, but lingering   
uncertainties on the HB theoretical brightness (due to the already   
discussed uncertainties on the $M_{core}^{He}$ value at the He-flash)   
and GC metallicity scale, leaves open the possibility that a   
discrepancy at the level of $\sim$0.20 mag between theory and   
observations may still exist, and one possibility to solve this  
discrepancy, if real,   
would be to consider overshooting from the convective envelope, as in the  
P00 models.   
   
\subsection{Stellar counts across the RGB Bump.}   
   
The brightness of the LF bump is a diagnostic of the maximum extension   
of the convective envelope reached at the base of the RGB, but it   
does not provide information on the size of the jump in the H profile   
left over when the envelope starts to recede.   
Since the evolutionary rate along the RGB is strongly affected by any change in the   
chemical profile, it is clear that this information can be obtained    
by analyzing the star counts in the bump region.   
This is a further test for the accuracy and adequacy of canonical RGB models   
which has been performed by Bono et al.~(2001).   
They have defined a new parameter -- $R_{bump}$ --   
which is the ratio between the number of stars in the region within $V_{bump}\pm0.4$ mag and the   
number of RGB stars in the interval $V_{bump}+0.5<V<V_{bump}-0.5$,    
(where $V_{bump}$ is the central magnitude of the bump region)   
and compared theory with observation for a large sample of GCs, using   
the theoretical models by CS97.   
   
Figure~\ref{rbumpteo} shows this comparison, and displays also the   
small dependence of $R_{bump}$ on age and initial He content.    
It is also evident that the selected [Fe/H] scale (Carretta \&   
Gratton~1997 in this case) and [$\alpha$/Fe] values (the same as in Fig.~\ref{bumpvhbb})   
are not critical, due to the small variation of the predicted    
$R_{bump}$ in the relevant metallicity range.   
Observational data are well reproduced by theory, with the few   
discrepant points being representative of clusters  with small RGB   
star samples and/or affected by differential reddening   
(see the discussion by Bono et al.~2001). Apart from these,
47Tuc (marked as an open circle in Fig.~\ref{rbumpteo})   
is the only cluster in clear disagreement with theory, in spite of the   
large sample of RGB stars observed, and deserves     
further detailed investigations.

\subsection{The RGB Bump as a diagnostic of partial mixing processes.}   
   
Cassisi, Salaris \& Bono~(2002) have recently shown that the RGB bump can be used   
to test not only the predicted size of the H-jump left over by the   
convective boundary at its maximum extension, but also the sharpness   
of this discontinuity. Canonical models and even models with   
overshooting -- as long as the mixing in the overshooting region is assumed to   
be instantaneous -- predict a sharp discontinuity of the H-profile.   
Conversely, any kind of non-instantaneous transport mechanism acting between the base of   
the convective envelope and a deeper region (where the local H abundance is  
no longer equal to the envelope value) which is able to build up   
abundance gradients, will smooth this sharp discontinuity, and affect   
the shape of the bump region in the LF (see also Bono \& Castellani~1992).   
   
Cassisi et al.~(2002) have computed models that simulate the   
occurrence of partial mixing processes at the bottom of the convective   
envelope of RGB stars. In particular, they have artificially modified   
the chemical element profiles of a model at the first dredge-up stage, smoothing out   
the sharp H-jump by means of a linear interpolation between the   
chemically homogeneous envelope and a region below the H-discontinuity.   
Different smoothing lengths   
-- parametrized as a function of the local pressure scale height --   
have been considered, and some representative results are shown in    
Fig.~\ref{smootlf}. The bump shape produced by canonical models    
appears asymmetric when the bin size is kept sufficiently small and    
photometric errors are neglected, with its characteristic peak located in the   
brightest bump region. By applying some smoothing of the chemical   
discontinuity, the variations in the shapes of   
the LF bump are quite evident: when the smoothing length increases the   
bump becomes more centrally peaked and    
more symmetric. This occurrence is a consequence of the changes in the   
evolutionary rate of the H-burning shell when it encounters the   
H-discontinuity (see Cassisi et al.~2002 for more details).   
Larger smoothing regions imply a smoother discontinuity in the   
H-profile (and in the molecular weight profile), and a   
smaller change in the efficiency of the H-burning shell.   
In particular, the narrowing of the bump width   
is related to the decrease of the luminosity drop occurring on the   
corresponding evolutionary track.    
   
The agreement between predicted and observed    
$R_{bump}$ values also puts constraints to the maximum possible extension   
of the partial mixing region, which increases the He abundance in the   
envelope, thus reducing the overall jump in the H abundance and   
modifying the value of $R_{bump}$. Smoothing lengths much   
larger than 1$H_{p}$ can be ruled out by the $R_{bump}$ test.   
Also the brightness level of the bump is   
modified by this process, becoming dimmer by   
$\approx0.025$ mag for a variation by $0.1H_p$ of the smoothing length.   
   
Data plotted in Fig.~\ref{smootlf} show that smoothing lengths equal to
or larger than $0.5H_p$   
affect significantly the shape of the LF bump.    
To check if this effect is detectable with real GC data,    
Cassisi et al.~(2002) have performed extensive Monte-Carlo simulations   
showing that in metal-rich GCs   
($[M/H]\approx-1.0$) not affected by differential reddening,    
the effect of smoothing lengths equal or larger than   
$0.5H_P$ can be detected with stellar samples larger than    
$\approx120$ objects within $\pm0.20$ mag of the bump peak, a bin size   
of at most 0.06 mag and   
$1\sigma$ photometric errors not larger than 0.03 mag (see Fig.~\ref{smootlf}).    
These requirements can be met, at least   
for the more massive, metal-rich GCs, by using the new   
Advanced Camera for Surveys (Clampin et al. 2000) on board the HST.    
   
\subsection{The RGB luminosity function heap}   
   
Ground based and HST   
observations of a large sample of RGB stars in NGC~2808    
reported by Bedin et al.~(2000), have led to    
the discovery of an additional bump in the LF of the GC NGC~2808 (see Fig.~\ref{heap}).    
This bump, termed \lq{heap}\rq, appears at visual magnitudes $\approx1.4$ mag brighter    
than the expected bump, and it is present both in the HST   
data -- containing the    
central regions of the cluster -- and in the ground-based data.    
{From} a preliminary analysis, this feature appears real and, apparently,   
lends support to the evidence    
that a similar heap is also present in the HST data for the   
cluster 47~Tuc (Bedin et al.~2000).    
   
Canonical RGB stellar models do not predict any decrease of the
evolutionary rate able to produce   
an increase in the star counts in the relevant region of the RGB. If   
the heap is assumed to be due   
to a change in the efficiency of H-burning, it could be interpreted as the byproduct of a   
deep-mixing episode,   
which somehow causes a temporary delay in the   
advancement of the H-shell.   
If this is the case, the LF heap could be adopted to constrain the extension of this   
deep-mixing in the stellar envelope. At the moment, one can conclude
that if the heap is produced by   
deep-mixing processes, canonical stellar models    
underestimate the evolutionary lifetime during this phase by about 25\%.   
   
However, Bedin et al. (2000) have noticed that, in the case of 47~Tuc,    
the position of the heap in the LF is similar to the location   
of the K Giant Variables (KGVs) discovered in this cluster by Edmonds \&   
Gilliland~(1996). If the connection between the LF   
heap and the presence of the KGVs were to be confirmed by further observations,   
the heap could be simply related to the position along the RGB    
of the KGV instability strip and to their properties during the pulsational cycle.    
   
   
\section{The TRGB as standard candle}   
   
As discussed in \S 3, the $I$-band magnitude of the TRGB appears to be a    
very good standard candle. Its predicted value   
depends on the TRGB bolometric magnitude derived   
from the models which, in turn, depends    
on the size of the He-core at the He-flash, and it is influenced    
by any uncertainty affecting the equation of state of partially electron degenerate   
matter, neutrino energy losses, electron conduction opacity,    
and the nuclear cross section for the   
$3\alpha$-reaction.    
   
   
   
In Fig.~\ref{comptip} we show   
a comparison of the most recent results concerning     
the TRGB bolometric magnitude and $M_{core}^{He}$ at the He-flash; the   
displayed quantities refer to a 0.8$M_{\odot}$ model and various   
initial metallicities (scaled solar metal distribution).   
$\rm M_{\rm bol}^{\rm TRGB}$ values have all been    
computed from log(L/$\rm L_{\odot}$) transformed   
into $\rm M_{\rm bol}$, assuming in each case $\rm M_{bol, \odot}$=4.75. This is   
equivalent of having a $BC_{V}$ scale in which $BC_{V, \odot}=-$0.07.  
  
There exists fair agreement among the various predictions   
of the $\rm M_{bol}^{\rm TRGB}$ metallicity dependence, and all 
the $\rm M_{bol}^{\rm TRGB}$ values    
at a given metallicity are in agreement within $\sim 0.10$ mag, with the exception    
of the P00 and YY01 models, which appear to be underluminous with   
respect to the others.    
As for the P00 models this difference follows from their smaller $M_{core}^{He}$ values;   
it is worth noticing that the recent models by Salasnich et   
al.~(2000), which are an update of the P00 ones, provide    
brighter $\rm M_{bol}^{\rm TRGB}$, similar to the V00 results. 
In case of the YY01 results the fainter TRGB luminosity cannot be explained
by much smaller $M_{core}^{He}$ values, since this quantity is 
very similar to V00 results.
   
If one neglects for a moment the P00 and YY01 models, the reason for the 0.1 mag   
spread among the rest of the authors is mainly due to the electron   
conduction opacities. The V00 (and Salasnich et al.~2000)    
models employ the HL69 data, while the other models use the I83   
result. Taking into account the effect of different choices for   
the opacities, the corresponding $M_{core}^{He}$ values would agree   
within 0.01$M_{\odot}$ and the difference in $\rm M_{bol}^{\rm TRGB}$ would further   
reduce .   
   
   
   
As a reference, we show here the relationship from the SC98 models   
(using $\rm M_{bol, \odot}$=4.75):   
   
\begin{equation}   
\rm M_{bol}^{\rm TRGB}=-3.949 - 0.178\rm [M/H] + 0.008\rm [M/H]^2   
\end{equation}   
   
This relationship covers the metallicity range   
$-2.35\le \rm [M/H]\le-0.28$, which encompasses    
the whole metallicity range spanned by galactic GCs.   
   
\subsection{Observational tests}   
   
Salaris \& Cassisi~(1997) and Caloi et al.~(1997)   
have discussed an important test for the consistency of theoretical TRGB   
brightness and He-core masses with observations.    
They have used the FPC83 estimates of   
the apparent bolometric TRGB magnitude of various GCs, obtained     
by directly integrating the flux received from their program stars via   
the observed UBVJHK photometry.    
After correcting these magnitudes   
for the GC distance moduli obtained   
using the distance scale provided by their    
ZAHB models, these authors have compared    
the theoretical $\rm M_{bol}^{\rm TRGB}$ values with the observed ones.   
This is essentially a comparison of the consistency between the   
theoretical TRGB and ZAHB distance scales; the importance    
of this test is due to the fact that the ZAHB and TRGB brightnesses    
depend in a different way on the value of   
$M_{core}^{He}$ and stellar metallicity.    
Consistency among them provides a strong    
indication of the reliability of the models.   
   
Figure~\ref{tip} shows a comparison between the FPC83, together with 
the more recent Ferraro et   
al.~(2000) data, and the theoretical relationship given above    
(which is on the same scale as the bolometric   
magnitude provided by the quoted papers), using the ZAHB distance   
scale given by the same SC98 models.   
The solid line represents the theoretical $\rm M_{bol}^{\rm TRGB}$ value,   
while the dashed lines mark $\rm M_{bol}^{\rm TRGB}$ values increased at steps of 0.1 mag,   
to show how the observational data are distributed in the various   
magnitude intervals.   
   
All observational points lie below the theoretical value, being on   
average about 0.10-0.20 mag underluminous. This is   
exactly what is expected in case of consistency between the ZAHB   
and TRGB distance scale, since, given the size of the RGB star samples   
observed by FPC83 and the evolutionary timescales close to the TRGB,   
it is statistically unlikely to have detected a star exactly at the   
TRGB (see, e.g., the discussions   
in Raffelt~1990, Castellani, Degl'Innocenti \& Luridiana~1993,    
Salaris \& Cassisi~1997). Caloi, D'Antona \& Mazzitelli~(1997) obtained   
the same results with their RGB and ZAHB models.   
   
This comparison enables us to perform another independent test    
for the efficiency of deep mixing in RGB stars (see \S 7).   
Sweigart~(1997a) has computed some evolutionary models for low-mass RGB   
stars considering mixing of some helium from the   
H-burning shell into the envelope. A consequence of this mixing   
is to increase both the TRGB and the ZAHB luminosity level.    
By parametrizing the efficiency of the deep mixing in term of the amount    
of hydrogen $\Delta{X_{mix}}$ carried into    
the H-shell, Sweigart found that $\log{L_{TRGB}}\approx0.8\Delta{X_{mix}}$ and    
$\log{L_{HB}}\approx1.5\Delta{X_{mix}}$. Since the HB is twice as   
sensitive as the the TRGB to the efficiency of this extra mixing process,    
one should expect that the comparison between ZAHB and TRGB distances    
discussed before could constrain the efficiency of this mechanism, if   
to occur at all. Assuming all stars in a given cluster experience   
this mixing process, for a given value of $\Delta{X_{mix}}$ the ZAHB brightness   
increases more than the TRGB one, thus moving upwards the TRGB   
absolute brightness with respect to the theoretical prediction.   
In the bottom panel of Fig.~\ref{tip}, we show this comparison   
in the case of $\Delta{X_{mix}}=0.10$. On average the   
observed TRGB magnitude is now equal to the theoretical one, showing 
that, on the basis of this test alone,   
$\Delta{X_{mix}}=0.10$ is a firm upper limit to the deep mixing    
efficiency (see also \S 7.3)  
\footnote{It is worth recalling that the same kind of test has been performed   
in other investigations, in order to put firm    
constraints on additional physical effects affecting    
the size of the He-core at the He-flash, such as the neutrino    
electromagnetic properties   
(Raffelt~1990, Raffelt \& Weiss~1992, Castellani \&   
Degl'Innocenti~1993), and extra spatial dimensions (Cassisi et al.~2000).}.   
   
\subsection{The TRGB absolute brightness calibration}   
   
$I-$band TRGB distance estimates are routinely obtained for    
Local Group galaxies, and the use of the HST capabilities has   
allowed the Leo I group and Virgo to be reached  
(e.g., Sakai et al. 1997, Harris et al. 1998).   
Madore \& Freedman~(1995) have discussed the observational  
requirements for obtaining reliable and unbiased   
TRGB detections.   
   
The absolute value of the TRGB distances depends, of course,    
on the absolute calibration of $\rm M_{bol}^{\rm TRGB}$ and the value of    
the bolometric correction to the $I$ band ($BC_{I}$). As previously discussed, the   
large majority of current RGB models predict $M_{bol}^{TRGB}$   
values showing a total dispersion of only $\sim$0.10 mag, and a dependence   
on the metallicity which is in fair agreement among various authors.   
To compare various theoretical determinations of the TRGB distance scale 
at a fixed metallicity, e.g., the average metallicity of the RGB
stellar population in the GC $\omega$~Cen
([Fe/H]$\approx -$1.7, see Bellazzini, Ferraro \& Pancino~2001 -- the
reason for this choice will be clear in the following),  
we will adopt the empirical $BC_{I}-(V-I)$ relationship   
by Da Costa \& Armandroff~(1990 -- which is on a scale where $BC_{V,   
\odot}=-$0.07), and the relationship between  $(V-I)_{0}^{TRGB}$ and
[Fe/H] from Bellazzini et al.~(2001). 
By assuming [$\alpha$/Fe]=0.3 and using Eq.~2,   
the various theoretical calibrations in fig.~\ref{comptip}, with the   
exception of the YY01 and P00 models, provide 
$M_{I}^{\rm TRGB}$ values ranging   
from $-$4.22 (Cassisi et al.~1998) to $-$4.10 (V00). 
The models of YY01 and P00 provide, respectively,   
$M_{I}^{TRGB}=-$4.00 and $-$3.95.  
As a reference, the $M_{bol}^{TRGB}$   
relationship by SC98 given before provides   
$\rm M_{I}^{\rm TRGB}=-$4.18, 
while the widely used semiempirical calibration by LFM93 gives 
$\rm M_{I}^{\rm TRGB}=-$4.00.  
This difference is easy to explain. The   
LFM93 relationship is based on a sample of clusters for which FPC83    
$\rm m_{bol}^{TRGB}$ data are available; 
the metallicity dependence is taken from   
theoretical models, while the zero point is based on the FPC83 $\rm
m_{bol}^{TRGB}$ values corrected for distance moduli provided by 
a HB brightness scale which is $\sim$0.1 mag fainter than the SC98
one. It is evident, on the basis of the   
previous discussion, that on average the FPC83 empirical data   
underestimate the true apparent TRGB brightness, and this occurrence accounts for another   
$\sim$0.1 mag of difference between the SC98 and LFM93 calibrations.
It is also worth mentioning that the use of theoretical bolometric   
corrections instead of the empirical Da Costa \& Armandroff~(1990) ones,
change the theoretical $M_{I}^{\rm TRGB}$ 
values by about $\pm$ 0.05 mag.
   
   
Very recently, Bellazzini et al.~(2001) have published an    
independent empirical recalibration of $\rm M_{I}^{\rm TRGB}$, based on the globular   
cluster $\omega$~Cen. They adopted the    
distance obtained by Thompson et al.~(2001) from the   
analysis of an eclipsing binary system in the cluster, and determined     
the TRGB apparent I brightness from a well populated CMD. The number   
of stars in the upper RGB appears to be large enough to provide   
an unbiased estimate of the TRGB level.    
{From} their assumed distance modulus and cluster reddening,   
$\rm M_{I}^{\rm TRGB}=-4.04\pm$0.12 is obtained. 
All of the theoretical results mentioned above (coupled with the empirical   
bolometric corrections by Da Costa \& Armandroff~1990)   
are within 1.5$\sigma$ of this calibration.   
   
\section{Conclusions}   
   
RGB theoretical models play a key role in the interpretation of 
various astrophysical observations. We have discussed in detail the 
theoretical uncertainties still affecting the models, paying particular attention 
to the predictions of colors, luminosities, evolutionary 
timescales and surface chemical abundances. A comparison of various theoretical 
RGB models with a large body of diverse empirical data has allowed us 
to discuss not only the accuracy of the 
adopted input physics, but also the adequacy of the assumptions built 
into canonical RGB models. 
 
Large uncertainties still exist in the predictions of $T_{\rm eff}$ 
and colors, due mainly to the treatment of superadiabatic convection, 
boundary conditions and color transformations employed; they  
can however be minimized by a suitable calibration 
of the models on empirical data. The prediction of the absolute 
value of the He-core mass at 
the He-flash -- which determines the TRGB and HB brightness -- suffers 
from residual uncertainties mainly related to the determination of 
accurate electron conduction opacities in the relevant physical regime. 
 
Extra mixing processes not included in canonical RGB models seem to 
be required to help explaining some of the chemical abundance patterns  
observed at the photosphere of RGB stars; however, independent empirical 
constraints, and arguments from stellar nucleosynthesis require that 
these processes should not be able to affect appreciably  
the evolutionary timescales, He-core masses and CMD location obtained 
from canonical models.

\begin{figure}  
\plotone{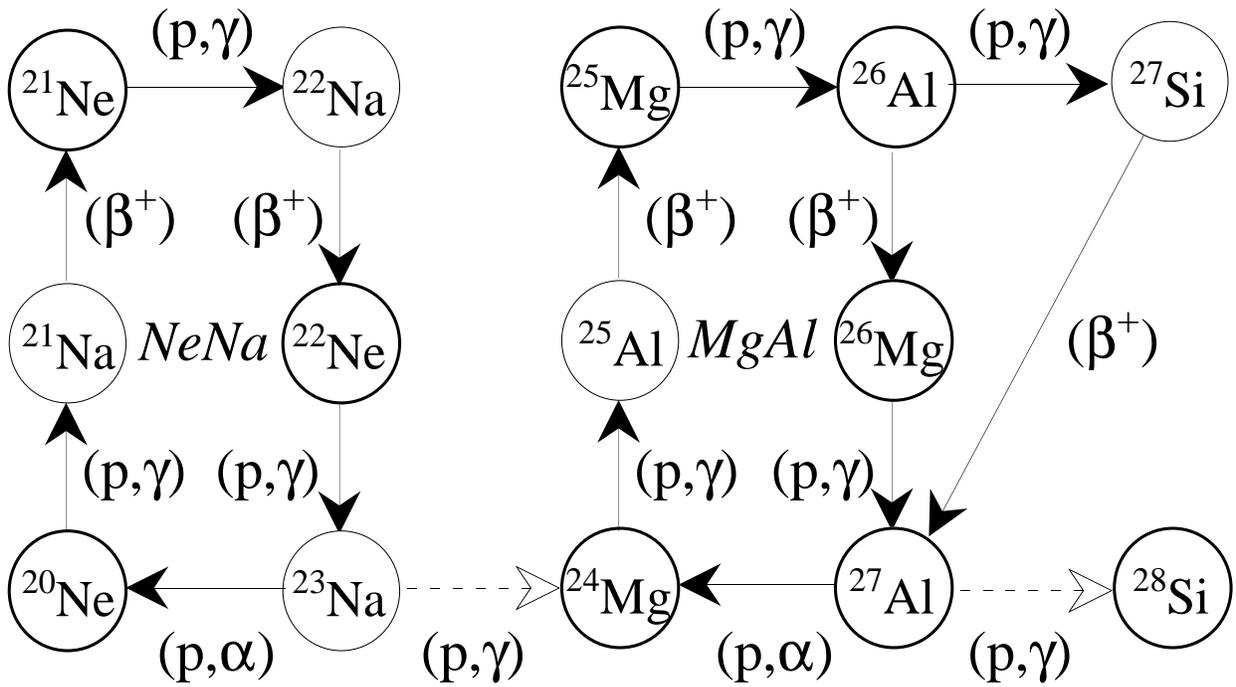}  
\caption{Reaction scheme of the NeNa- and MgAl-proton-capture  
cycles. The two reaction sequences leading to the creation of an
$\alpha$-particle are very similar to the  
CNO-I-cycle. (Figure courtesy of S.~Goriely) \label{f1}}   
\end{figure}

\begin{figure}  
\plotone{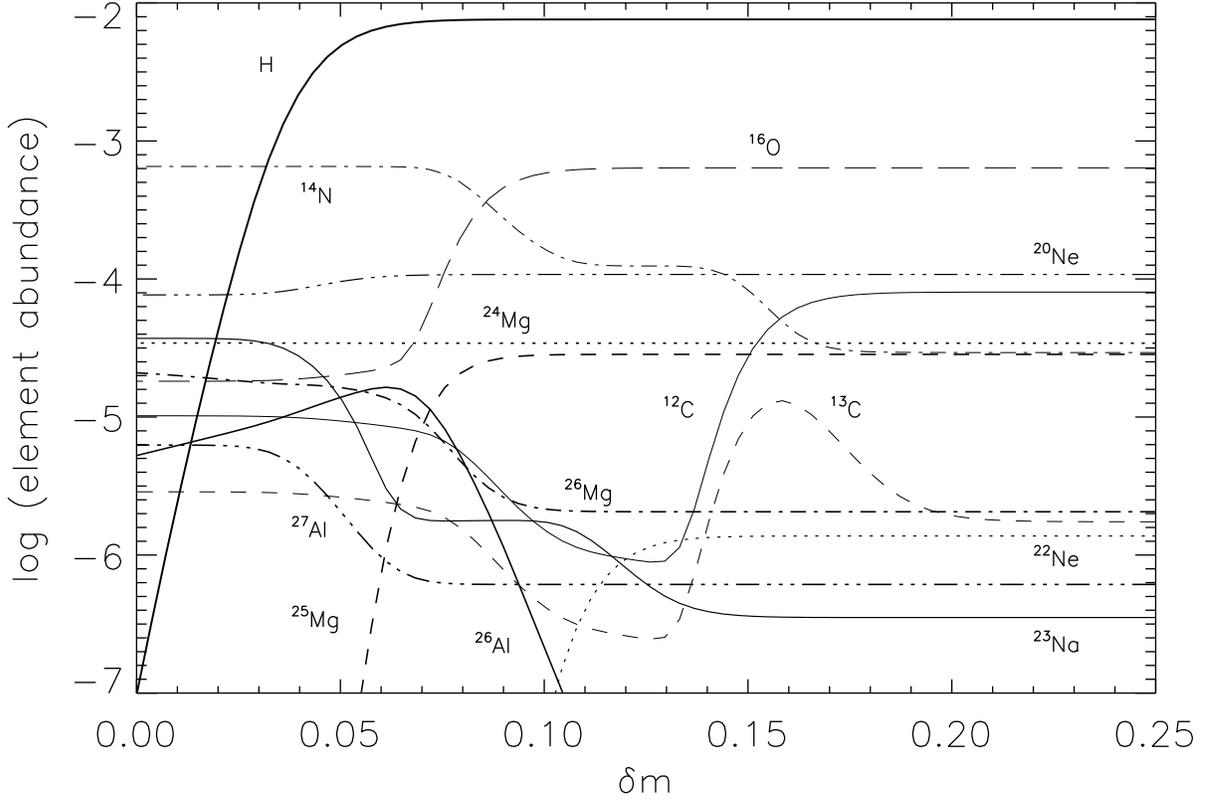}  
\caption{Abundance profiles (logarithmic mass fractions) within the hydrogen shell of an  
$0.8\, M_\odot$ RGB-model of metallicity $\mathrm{[Fe/H]}=-1.58$. For  
this model, taken from Denissenkov \& Weiss~(2001),  
$\mathrm{[^{25}Mg/Fe]}=1.2$ was assumed (primordial enrichment) and the standard  
NACRE-rates were used. The abscissa is in a relative mass  
coordinate $\delta m$, which is defined as 0 at the bottom of the H-shell and 1  
at the bottom of the convective envelope. Plotted are the most important  
isotopes of the three proton-capture cycles, which are grouped by  
line thickness. Hydrogen has been scaled down by 1/100 to fit on the  
same scale. $^{26}$Al refers to the ground state.  
\label{f2}}  
\end{figure}

\begin{figure}  
\plotone{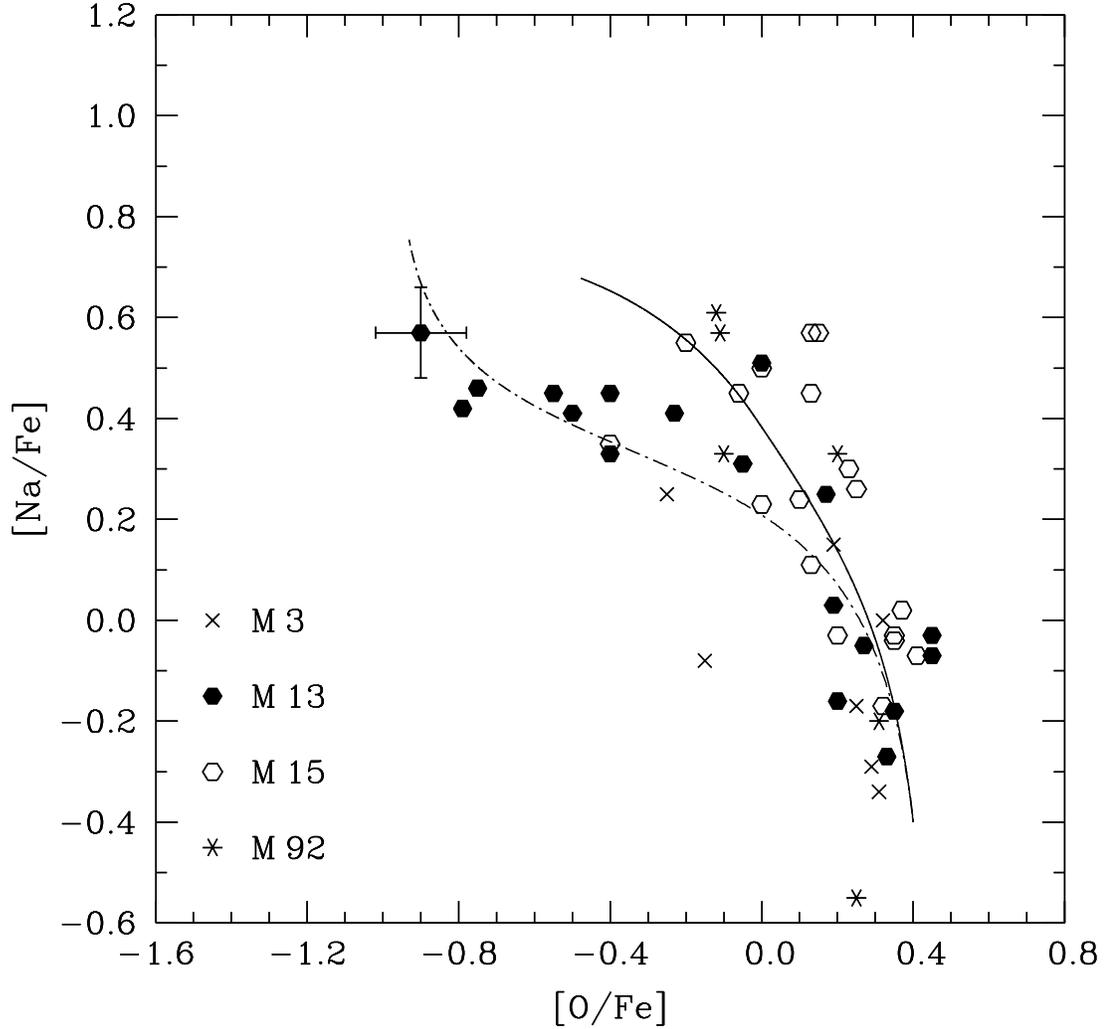}  
\vspace{-2.5cm}  
\caption{Observed Na-O-anticorrelation in a number of clusters  
(symbols; one with typical error bars), along  
with  model predictions: the dot-dashed line is for  
mixing parameters of $2.5\cdot  
10^9\,\mathrm{cm}^2\mathrm{s}^{-1}$ for the constant of diffusive  
mixing and $\delta m = 0.06$ for penetration depth (Denissenkov et al.~1998); the  
solid line is   
for rotation-induced mixing following the model described in  
Denissenkov \& Tout~(2001) with an angular velocity at the bottom of the convective  
envelope of $8\cdot 10^{-6}\,\mathrm{rad\,s}^{-1}$. Note that along the  
lines time increases as a parameter, while the data points are not  
necessarily ordered this way.  
\label{f3}}  
\end{figure}

\begin{figure}  
\plotone{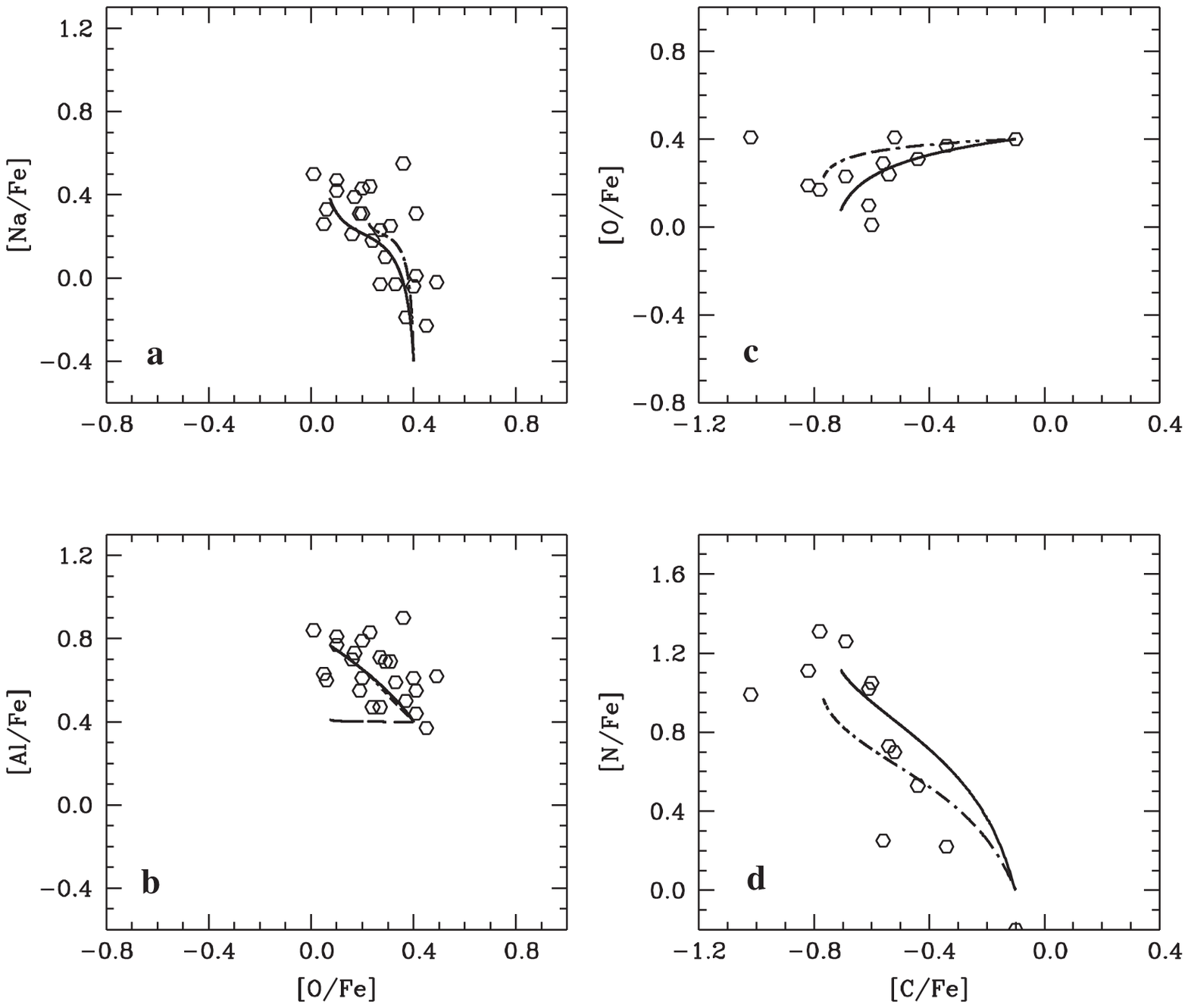}  
\caption{Observed element correlations in M4 (data by  
Ivans et al.~1999) and models (lines) by Denissenkov \& Weiss~(2001). The solid and  
dot-dashed lines correspond to two different penetration depths,  
$\delta m = 0.065$ and $0.075$, respectively; the diffusive constant  
was $4\cdot 10^8\,\mathrm{cm}^2\mathrm{s}^{-1}$ in both cases; the  
dashed line in panel b refers to the evolution of the stable  
$\Iso{27}{Al}$ isotope for the first parameter set; it  
clearly fails to reproduce the data. The other lines in this panel  
show the $\Iso{26}{Al}^{\rm g}$   
abundance. For details and a similar plot  
for $\omega$~Cen see Denissenkov \& Weiss~(2001).\label{f4}}  
\end{figure}

\begin{figure}   
\plotone{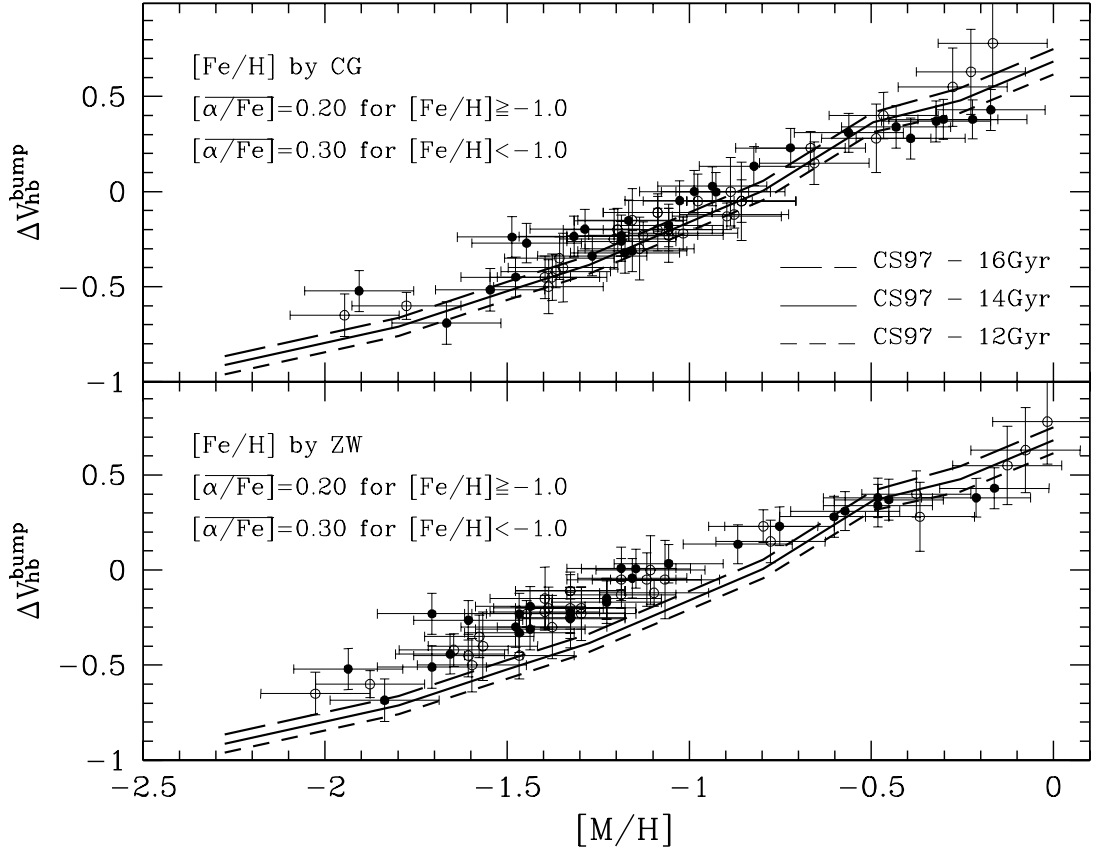}   
\caption{Comparison between the theoretical (CS97 models)   
and empirical values of \vhbb as a function   
of [M/H]. Full circles refer to the data by Zoccali et al.~(1999), empty   
circles to the data by Ferraro et al.~(1999). [M/H]   
follows the Carretta \& Gratton~(1997) 
[Fe/H] scale together with the labelled assumption   
about the $\alpha$-enhancement (top panel). The bottom panel shows the   
same data but according to the Zinn \& West~(1984) [Fe/H] scale.\label{bumpvhbb}}    
\end{figure}

\begin{figure}   
\plotone{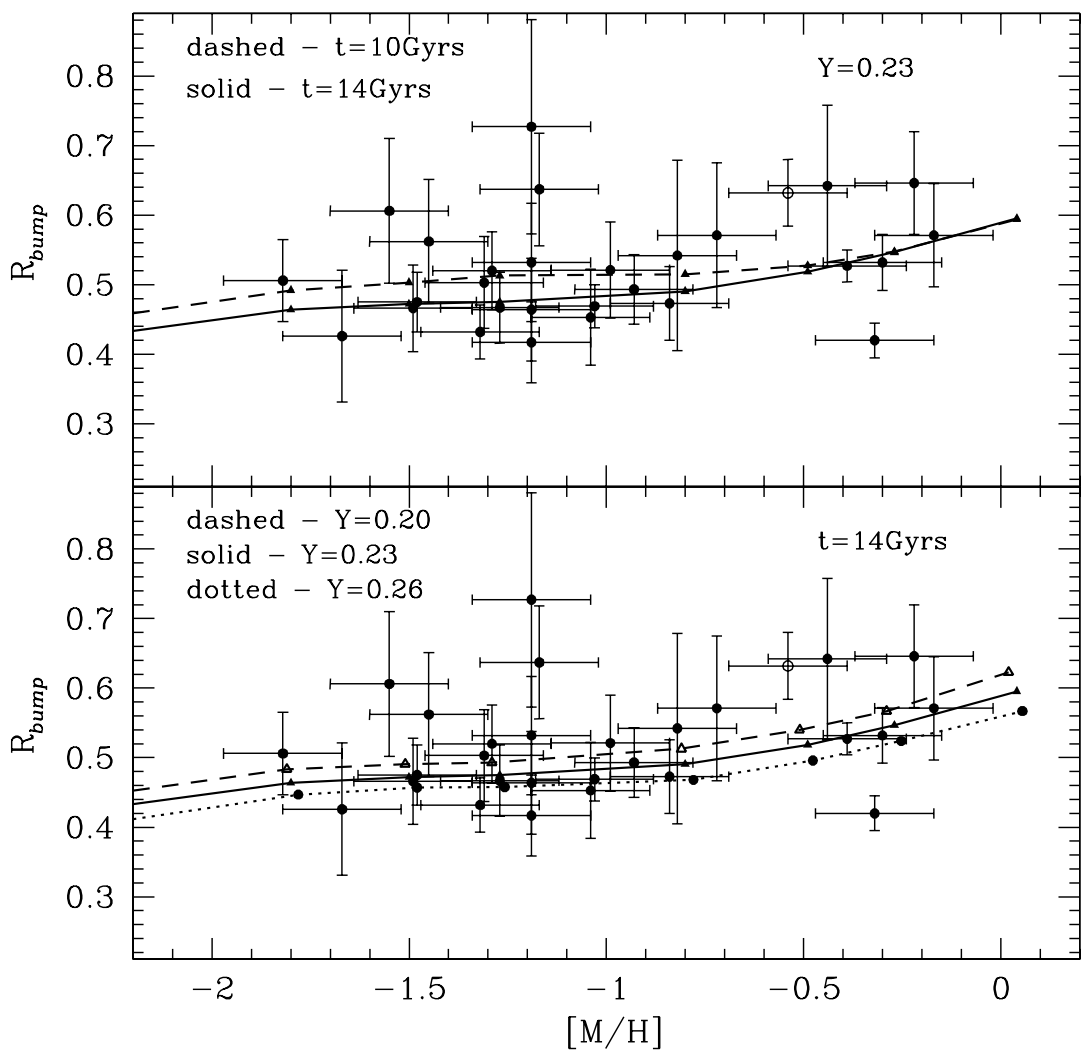}   
\caption{Comparison between observed (from Bono et al.~2001) and predicted (CS97 models)   
values of the $R_{bump}$ parameter. The top panel shows the effect of   
varying age on the theoretical $R_{bump}$ values; the bottom panel the effect    
of varying the initial He abundance. The open circle refers to the GC 47Tuc.\label{rbumpteo}}    
\end{figure}

\begin{figure}   
\plotone{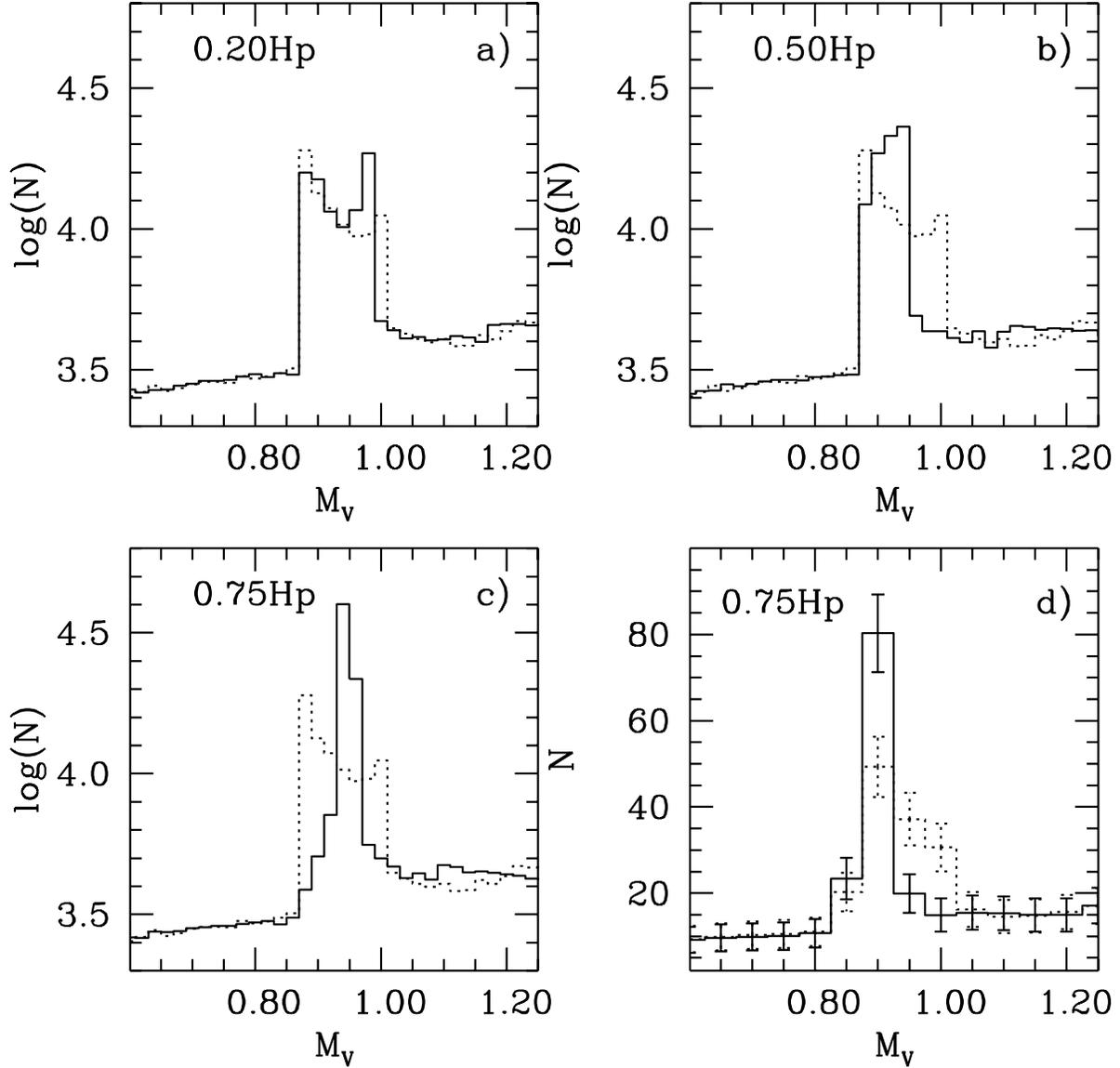}   
\caption{Panels a), b) and c) show a comparison between the LF from CS97 canonical models (dotted   
line) of 1$M_{\odot}$ and Z=0.006, and LFs obtained   
accounting for various degrees of smoothing (as labelled) of the 
H-jump (solid lines). The bin size is 0.02 mag and all the LFs are 
normalized to the same arbitrary number of stars above the bump region. 
Panel d) shows the same comparison in the case of a smoothing of  
0.75$H_{p}$, but for the case of Monte-Carlo simulations with 200 
stars within $\pm$0.20 mag of the bump peak, 
bin size of 0.05 mag and 1$\sigma$ photometric errors of 0.015 mag;  
1$\sigma$ error bars on the stellar counts are also displayed. 
In each panel the LFs of the  non-canonical models have been shifted 
in luminosity in order to match the bright   
end of the bump obtained from canonical models.\label{smootlf}}    
\end{figure}

\begin{figure}   
\plotone{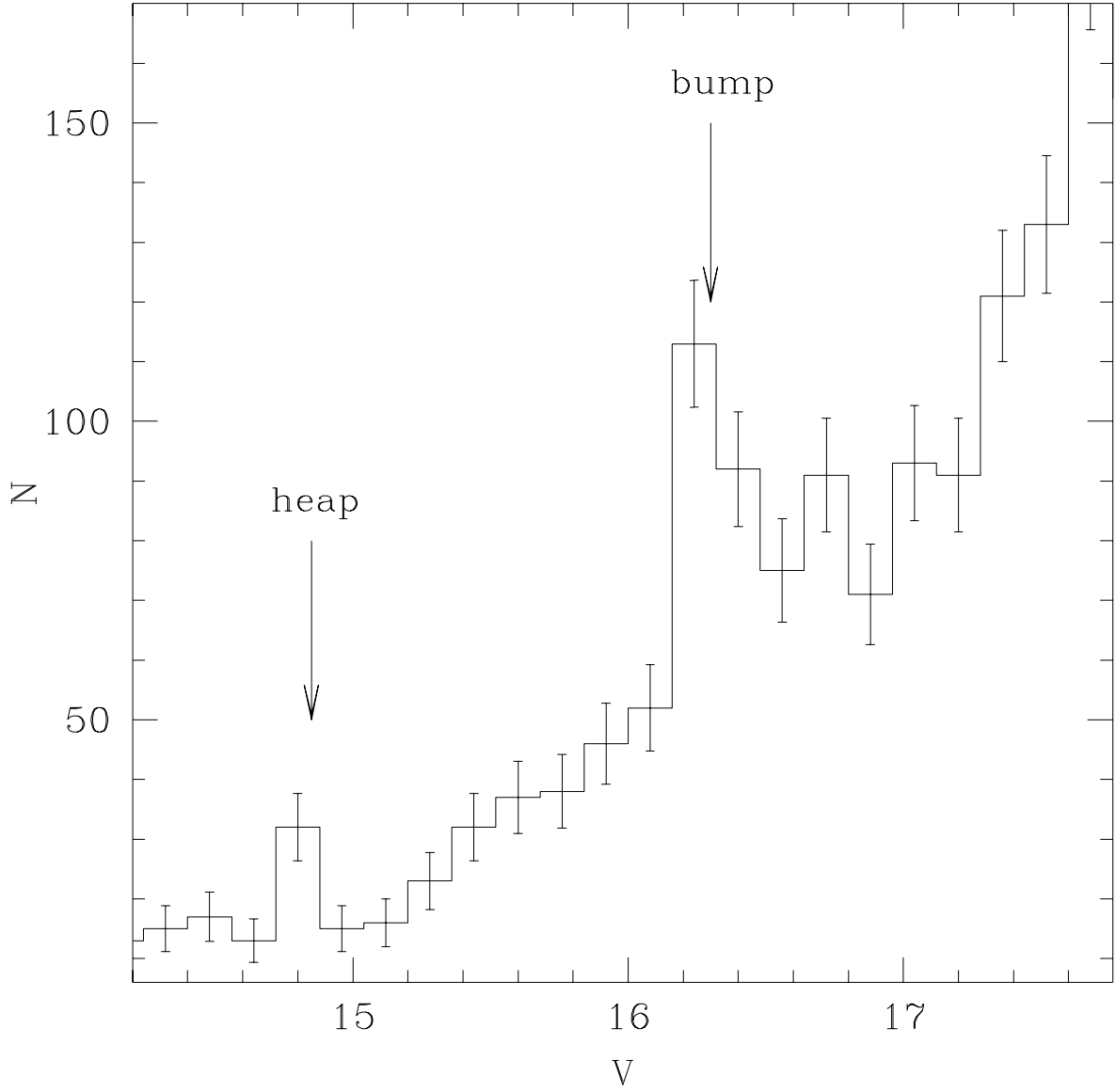}   
\caption{The LF of the RGB of NGC~2808. The arrows mark the position of the bump and   
the newly discovered heap.   
\label{heap}}    
\end{figure}

\begin{figure}   
\plotone{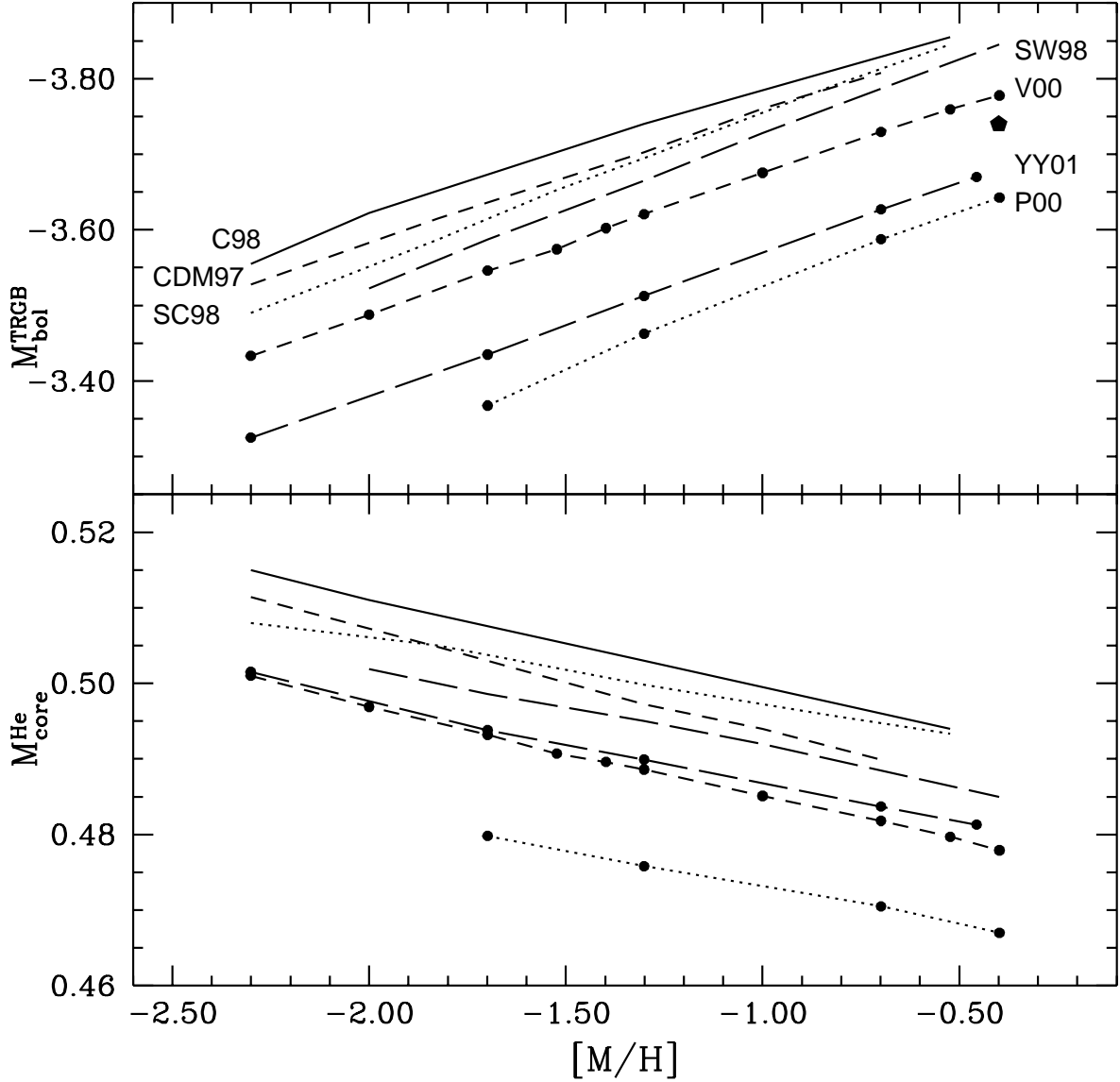}   
\caption{$M_{bol}^{TRGB}$-[M/H] and $\rm M_{core}^{He}$-[M/H] (in solar mass   
units) relationships    
for 0.8$M_{\odot}$ models from different authors. The meaning of the   
labels is as in Fig.~\ref{rgbcomp}, with the addition of   
Cassisi et al.~(1998) -- C98; Caloi,   
D'Antona \& Mazzitelli~(1997) -- CDM97; Salaris \& Cassisi~(1998) --   
SC98. The filled symbol at [M/H]=$-$0.4 corresponds to the models by Salasnich   
et al.~(2000).   
\label{comptip}}    
\end{figure}

\begin{figure}   
\plotone{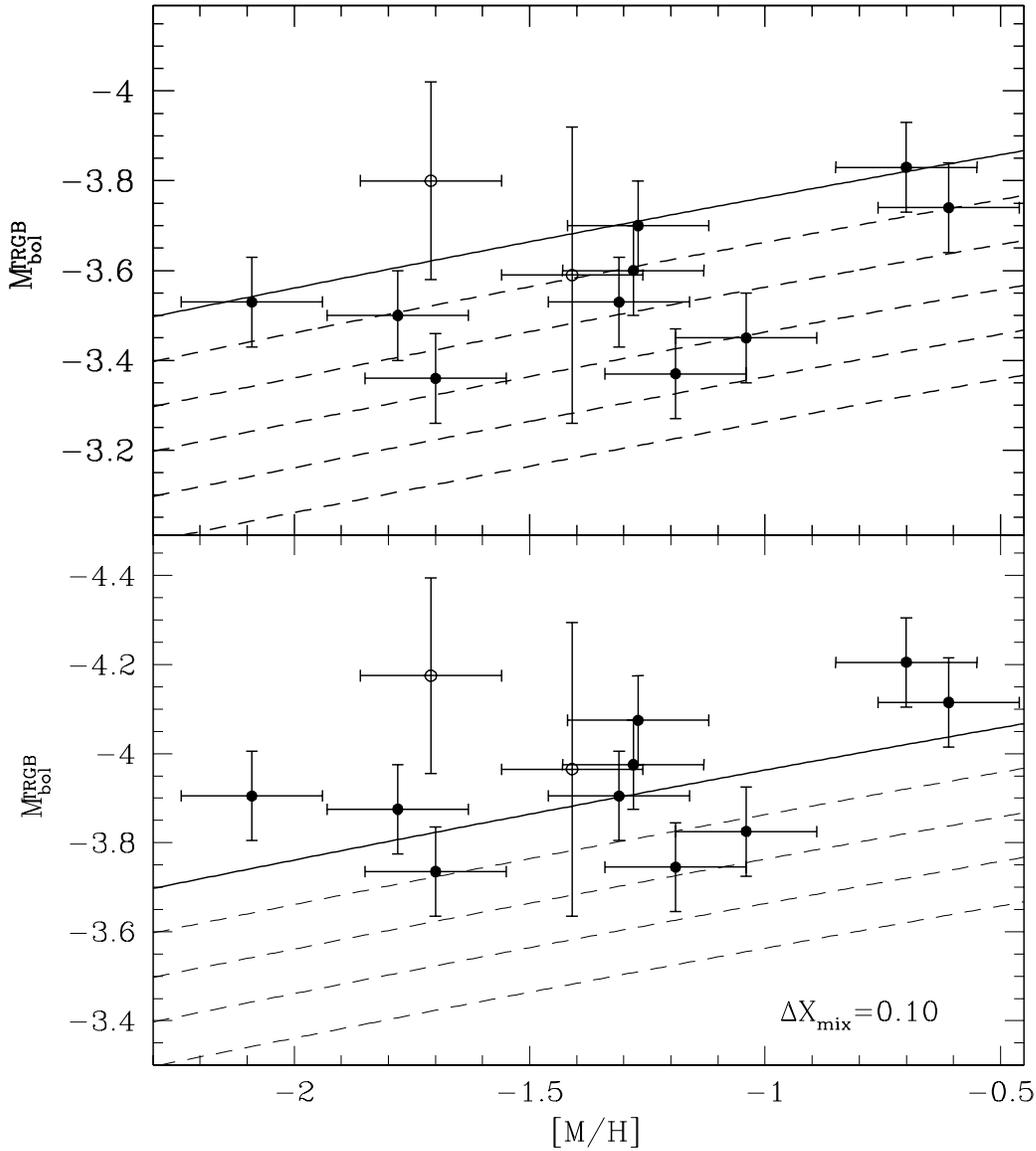}   
\caption{Comparison between the absolute bolometric magnitude of the brightest   
RGB star observed in selected GCs by FPC83 (full circles) and by   
Ferraro et al. (2000 --   
empty circles) corrected for the distance moduli obtained from the   
ZAHB models by CS98, and theoretical TRGB results by CS98 (solid line).    
The dashed lines represent the same theoretical relation but shifted   
in steps of 0.10 mag (top panel). The same is shown in bottom panel,     
but the theoretical TRGB value  and the ZAHB distance scale account   
for an efficiency of envelope He mixing equal to   
$\Delta{X_{mix}}=0.10$ (see text for details). \label{tip}}   
\end{figure}

\acknowledgments{ 
We wish to warmly thank V.~Castellani and D.~Vandenberg for 
very useful comments on an early version of the manuscript, 
which greatly improved the quality of this review, as well as
for valuable suggestions and discussions on the subject.
We also thank G.~Bono, L.~Girardi, P.~James, S.~Percival and L.~Piersanti
for an accurate reading of the manuscript.
We are grateful to  F.~D'Antona, D.~Vandenberg and S.~Yi
for providing us with their results, and to M.~Bellazzini, B.~Chaboyer, 
H.~Schlattl for useful discussions.
 
A.W. likes to thank P.~Denissenkov for his collaboration, R.~Kraft and  
R.~Peterson for many stimulating discussions and advices, and  
S.~Goriely for providing Fig.~\ref{f1}.   
S.C. acknowledges the financial support by 
MURST (Cofin2000) under the scientific project: ``Stellar observables 
of cosmological relevance'' and by (Cofin2001) under the scientific project:
``Origin and evolution of stellar populations in the galactic spheroid''.
}

\pagebreak    
   
\clearpage    
  
\clearpage  
   
\clearpage  
   
   
   
  
\clearpage

\clearpage

\clearpage  
  
\clearpage

\clearpage  
   
\clearpage  
   
\clearpage  
  
\clearpage  
  
\clearpage  
  
\clearpage  
  
\clearpage   
  

\clearpage  
  
   
\clearpage  
  

  
\clearpage  
  
\clearpage  
     
\end{document}